\documentstyle[11pt]{article}
\date{}

\begin{document}

\setcounter{page}{0}
\thispagestyle{empty}

\begin{flushright}
CERN-TH. 95-283
\end{flushright}
\vspace{0.2cm}

\begin{center}
{\Large
{\bf SPACE, TIME AND COLOR IN HADRON
\medskip

PRODUCTION VIA {\boldmath $e^+e^- \rightarrow Z^0$} AND
{\boldmath $e^+e^- \rightarrow W^+W^-$}
}
}
\end{center}
\bigskip

\begin{center}

{\Large
{\bf John Ellis and Klaus Geiger}
}

{\it CERN TH-Division, CH-1211 Geneva 23, Switzerland}
\end{center}
\vspace{1.0cm}

\begin{center}
{\large {\bf Abstract}}
\end{center}
\smallskip

The time-evolution of jets in hadronic $e^+e^-$ events
at LEP is investigated in both position- and momentum-space,
with emphasis on effects due to color flow and particle correlations.
We address dynamical aspects of the four
simultanously-evolving, cross-talking parton cascades that appear in
the reaction $e^+e^-\rightarrow \gamma^\ast / Z^0 \rightarrow W^+W^-
\rightarrow q_1\bar{q}_2q_3\bar{q}_4$,
and compare with the familiar two-parton cascades in
$e^+e^-  \rightarrow  Z^0 \rightarrow  q_1 \bar{q}_2$.
We use a QCD statistical transport approach, in which
the multiparticle final state is treated as an evolving mixture
of partons and hadrons, whose proportions are controlled
by their local space-time geography via standard perturbative
QCD parton shower evolution and a phenomenological model for
non-perturbative parton-cluster formation followed by cluster decays
into hadrons. Our numerical simulations exhibit
a characteristic `inside-outside' evolution
simultanously in position and momentum space.
We compare three different model treatments of color flow, and find
large effects due to cluster formation by the combination of
partons from different $W$ parents.
In particular, we find in our preferred model a shift of
several hundred MeV in the apparent mass of the $W$, which is
considerably larger than in previous model calculations.
This suggests that the determination of the $W$ mass at LEP2
may turn out to be a sensitive probe of spatial correlations
and hadronization dynamics.

\noindent

\vspace{0.5cm}

\rightline{johne@cernvm.cern.ch}
\rightline{klaus@surya11.cern.ch}
\leftline{CERN-TH. 95-283, October 1995}

\newpage

\noindent {\bf 1. INTRODUCTION}
\bigskip

One of the primary physics topics at the second phase of the
CERN $e^+e^-$ collider LEP is expected to be the determination
of the $W^{\pm}$ mass. Because of the higher statistics and
the possibility of using more kinematic constraints in
the absence of energetic neutrinos from leptonic $W^{\pm}$
decays, there is great interest in using purely hadronic
decays of $W^+W^-$ pairs. A major challenge in the analysis
of these events is the assignment of the observed hadrons to
the parent $W^{\pm}$, which may alternatively be regarded as
a novel opportunity to study `non-trivial' space, time and
color correlations of multiple partons
in the production and decay of $W^+W^-$ pairs.
Compared with jet fragmentation in $e^+e^-$ collisions
on the $Z^0$ peak, the expected new physics aspects
are due to the fact that the $W^+$ and $W^-$ decay
very rapidly after their production into two jets each, while they
are still practically on top of one other. Hence, one must deal
with an initial parton configuration consisting
of two quark-antiquark pairs that
are produced very close in both space and time.
The subsequent evolution of the system via parton showering and
parton-hadron conversion is therefore subject to interference
and combination between the decay products of the $W^{\pm}$, and
correlation effects due to the local density of particles.
This situation is to be contrasted with the more familiar
two-jet events, in which a single back-to-back initial
pair of partons evolves essentially unscathed in empty space.

Treating these interference, combination
 and correlation effects certainly requires a detailed
understanding of color structure in the development of hadronic
showers. This has become available during the past few years,
as analytical perturbative QCD calculations
within the modified leading-logarithmic approximation
\cite{MLLA} have matured to provide
a quantitative description of color flow phenomena in
jet events. Moreover,
the effect of the calculated perturbative color structure
on the non-perturbative hadronization process
has been considered within the local parton-hadron duality
framework \cite{LPHD}. The results have included
detailed predictions for many hadronic observables, which have been
found to agree remarkably well with experimental data
from LEP1 experiments on the $Z^0$ peak \cite{lep}.

The prospective scenario of cross-talking jets and the
resulting complications associated with multi-particle evolution
has, moreover, recently attracted considerable attention to
color correlation and interference effects associated with
the randomization and statistical rearrangement of the color flow
in $W^+W^-$ events at LEP2.
Perturbative aspects of this physics
 have been studied in a number of papers
by the Durham-St. Petersburg group \cite{khoze92,khoze93,khoze94},
and phenomenological non-perturbative aspects have
been investigated by various authors in Refs.
\cite{GPZ,SK,GH,BW,LL,sharka,GPW}.
In particular,
Sj\"ostrand and Khoze \cite{SK} have discussed in great detail
interference and color reconnection effects in overlapping
jet configurations during both the perturbative QCD regime of
parton evolution, where they are suppressed by $O(\alpha_s^2/N_c^2)$,
and the non-perturbative conversion of partons into final-state hadrons.
The perturbative analysis is firmly based
on the fundamentals of QCD, but in our present state of ignorance
the non-perturbative aspects
require a phenomenological description of the hadronization process.
In Ref. \cite{SK} this was addressed within the Lund
string-fragmentation model \cite{lund}, whereas in Ref. \cite{BW}
the Marchesini-Webber model \cite{herwig} was employed.
A common feature of both approaches was the use of a
particular model for the
rearrangement of color in hadronization,
which is modelled as a ``color reconnection'' of final-state partons,
in which some {\it ad hoc} elements are required
(see Ref. \cite{BW} for an overview of such color reconnection models).

These previous works have reached the important conclusions that,
on the one hand, perturbative
color rearrangement and interference effects are negligibly small,
but, on the other hand, final-state interactions in the
process of hadron formation may lead to a significant systematic
shift in the mass of the $W^{\pm}$ which could have
far-reaching implications for its determination at LEP2.
Since parton-hadron transmutation is currently not calculable from
first principles, but requires model assumptions,
the latter conclusion is necessarily model-dependent,
and a comparison of different approaches is therefore
desirable.

We think it may be useful to complement previous approaches
by considering the full space-time history of the
dynamically-evolving particle system, so that
the correlations of partons in space and time can be taken fully
into account when hadronization is modelled.
With this motivation in mind, we develop two new features
in the present work:
\begin{description}
\item[(i)]
we provide and test an alternative description of the interplay between
parton evolution and hadronization, in which we trace
the microscopic history of the particle dynamics in both space-time
and momentum space, and
\item[(ii)]
we study the possible observable effects of multi-parton correlations in
hadronic $e^+e^-$-events using a different space-time model of the
parton-hadron conversion process.
\end{description}
These features enable us to compare
the dynamics of ordinary 2-jet events with the evolution of
4-jet events that consist of two overlapping and
simultanously-evolving 2-jets in a way that differs from
previous approaches, and thus may cast useful light on the
possible model-dependence of their results.

The central element in our approach is the use of
QCD transport theory \cite{msrep} and quantum field kinetics \cite{ms39}
to follow the evolution of the multi-jet system
in 7-dimensional phase-space $d^3 r d^3 k dk^0$.
We include both the perturbative QCD parton-shower development
\cite{MLLA,jetcalc, bassetto},
and the phenomenological
parton-hadron conversion model which we proposed recently \cite{ms37},
considering dynamical parton-cluster formation as
primarily dependent on spatial separation, which is
followed by the decay of clusters into hadrons.
This approach may be extended to situations involving more
complicated initial states, as in $e$-$p$, $p$-$p$, $e$-nucleus,
$p$-nucleus and nucleus-nucleus collisions. However, the fact that
the initial state is unambiguous in $e^+e^-$ collisions provides
a particularly clean laboratory for studying the differences
between our approach and others, which is also of great topical
relevance. Specifically, in this paper we compare
\begin{equation}
e^+e^- \; \rightarrow \;  Z^0 \;\rightarrow \;
q_1 \bar{q}_2 \;\rightarrow \;hadrons
\label{ee1}
\end{equation}
at a center-of-mass energy of 91 GeV as at LEP1
with
\begin{equation}
e^+e^- \; \rightarrow \;  \gamma^\ast / Z^0 \;\rightarrow \;W^+W^-\;\rightarrow
\;q_1 \bar{q}_2 q_3 \bar{q}_4\;\rightarrow \;hadrons
\label{ee2}
\end{equation}
at a center-of-mass energy of 170 GeV as at LEP2.
The theoretical description of both the parton-shower evolution and
the hadronization of the produced quarks and gluons should be of the
same generic pattern in these two cases, namely
time-like parton cascades with hadron formation
in vacuum, i.e., in the absence of beam remnants or surrounding matter.
This enables key aspects of our approach to be isolated and compared
with previous approaches.
Since QCD at LEP1 is nowadays very well understood and described
impressively accurately \cite{MC} by various Monte Carlo models
\cite{pythia,herwigmc,ariadne}, the reaction
(\ref{ee1}) provides a testing ground for our distinct approach, as well
as a baseline for extracting new physics aspects
in the reaction (\ref{ee2}) at LEP2.

This paper is organized as follows.
In Sec. 2 we review our transport-theoretical approach to the space-time
description of general high-energy QCD processes,
and describe different ways of incorporating
color degrees of freedom, a feature we had not addressed previously.
To assess the model-dependence of the effects of
color flow on the space-time evolution, we define 3 distinct scenarios.
One does not consider color at all, as in our previous work, in a
second scenario
the color flow is traced so that only color-singlet
configurations of two partons participate in hadronization, and in the
third variant arbitrary color configurations participate in
hadronization, with
hadrons emerging from color-singlet components and the initial
net color being locally balanced by parton emission.
Sec. 3 is then devoted to the application and  systematic
analysis of this formalism to hadronic $e^+e^-$ events
of the types (\ref{ee1}) and (\ref{ee2}).
We demonstrate the emergence of the
inside-out cascade picture \cite{bj,kogut73} in these reactions,
and compare the above three scenarios for taking account of
color degrees of freedom,
confronting characteristic event properties and observables in the
three cases.
We target special attention on the $W$ mass determination in
4-jet events (\ref{ee2}), simulating the
experimental reconstruction of jets
and exploring how the apparent $W$ mass may be shifted by the
previous space-time history of the system,
in particular by interference between the parton showers
produced in the decays of the two different $W$ particles.
Finally, in Sec. 4 we discuss the interpretation
of our results, which suggest that the $W$ mass shift may be
even larger than that found in previous studies \cite{SK,BW}.
Much of this substantial difference in the magnitude
of particle- and color-correlation effects from the
previous investigations of Refs. \cite{SK,BW} may be due to the
fact that our hadronization Ansatz allows partons to form clusters
`exogamously' without any regard to their origins in the two perturbative
$W$ showers, the sole criterion being their spatial separations.
This aspect is significant in the particular space-time geography
of $W^+W^-$ events (\ref{ee2}), in which
the wee partons from the two parton showers overlap strongly
in the neighborhood of the
$W^+W^-$ production vertex.
\bigskip
\bigskip

\noindent {\bf 2. KINETIC APPROACH TO PARTON EVOLUTION AND HADRONIZATION}
\bigskip

In the first part of this section, we review essential elements of
the kinetic approach \cite{ms37} to the dynamical description
of perturbative QCD shower development and
parton-hadron conversion that we introduced previously, which
incorporates spatial information in an essential way.
Then we provide a first discussion of color flow within this
framework, proposing three scenarios for the treatment of color
degrees of freedom in cluster formation.
In the last part of this Section,
we outline the practical calculation scheme that we use in Section 3
as the basis for a Monte Carlo simulation of the space-time development
of $e^+e^-$-collisions.
\medskip

\noindent {\bf 2.1 Review of our Model}
\smallskip

In a previous paper \cite{ms37} we presented in detail our
approach to the problem of hadronization  of quarks and gluons
in conjunction with the space-time evolution of parton showers
produced by hard QCD processes in high-energy reactions, such as
$e^+e^-$ collisions, deep-inelastic $ep$ scattering, hadron-hadron
and eventually nucleus-nucleus collisions.
Here we just sketch the main
features of our model, and refer the
interested reader to our previous paper for more details.

In order to model the dynamical transition
between the perturbative and non-perturbative domains of QCD,
we advocate a combination of perturbative QCD and effective
field theory in their respective domains of validity,
and describe the conversion between them
using ideas developed in phenomenological
descriptions of the finite-temperature transition
from a quark-gluon plasma to a hadronic phase \cite{CEO}.
The latter is described by an effective theory incorporating
a chiral field $U$ whose vacuum expectation value ($vev$)
$U_0 \equiv \langle 0 |U+U^\dagger|0\rangle$
represents the non-perturbative
quark condensate $\langle 0 |\bar q q|0\rangle$,
and a scalar field $\chi$ whose $vev$ $\chi_0 \equiv
\langle 0 | \chi| 0 \rangle$
represents the non-perturbative gluon condensate
$\langle 0 | F_{\mu\nu}F^{\mu\nu}| 0 \rangle$.
We visualize hard scattering in high-energy processes
as producing a ``hot spot'', in which the usual long-range order
represented by $U_0$ and $\chi_0$ is disrupted locally by the
appearance of a bubble of the naive perturbative
vacuum in which
$\langle 0 |\bar q q|0\rangle = 0=
\langle 0 | F_{\mu\nu}F^{\mu\nu}| 0 \rangle$.
Within this bubble, a parton shower develops
in the usual perturbative way, with the hot spot expanding
and cooling in an irregular stochastic manner described by QCD
transport equations \cite{ms39}.
This perturbative description remains appropriate in any
phase-space region of the shower where the local energy density is large
compared to the difference in energy density between
the perturbative partonic and the non-perturbative hadronic vacua.
When this condition is no longer satisfied, a hadronic bubble
\footnote{
Specifically, this bubble has quantum numbers matching those of the
parent partons, and contains both gluonic ($\chi$) and chiral ($U$)
degrees of freedom.
We acknowledge critical questions from T. Sj\"ostrand on this point.
}
may be formed with a probability determined by
statistical-mechanical considerations.
A complete description of this conversion
requires a treatment combining partonic and hadronic degress of freedom,
which is an essential aspect of our approach.
\medskip

{\bf 2.1.1 Scale-Dependent Effective Lagrangian}
\smallskip

The theoretical basis for the above
intuitive picture is provided by
a gauge-invariant effective Lagrangian
that embodies both fundamental partonic degrees and effective
hadronic degrees of freedom.
It depends explicitly on the space-time scale $L$ (specified below)
which characterizes locally the relevant dynamical processes of the
physical system under consideration. As we shall see later,
the appropriate space-time scale $L$ in any region of a hadronic
shower is governed locally by
the dynamical evolution of the system itself, through kinetic
equations that are constrained by the uncertainty principle,
and incorporate both space-time and energy-momentum variables.
Our Lorentz-invariant measure $L$ of
the space-time separation between two color charges located
at $r_i$ and $r_j$ where $r\equiv r^\mu=(t,\vec r)$ is defined by
\begin{equation}
L(r_i,r_j)\;:=\;\sqrt{(r_i-r_j)_\mu(r_i-r_j)^\mu}
\;,
\label{Ldef}
\end{equation}
We introduce phenomenologically a confinement length-scale $L_c$
\begin{equation}
L_c \;= \;O(1 \; fm)
\;\;,\;\;\;\;\;\;\;\;\;\;
\alpha_s (L_c^{-2}) \;
= O(1)
\label{Lc}
\end{equation}
that characterizes the distinction between
short-distance ($L\ll L_c$)
and long-range ($L\,\lower3pt\hbox{$\buildrel > \over\sim$}\,L_c$)
physics in  QCD.
In the limit of short distances $L\ll L_c$,
where $\alpha_s(L^{-2})\ll 1$,
QCD is well described perturbatively by the usual fundamental Lagrangian
${\cal L}_L[A^\mu,\psi,\overline{\psi}]$
in terms of the elementary gluon ($A^\mu$) and quark fields
 ($\psi,\overline{\psi}$).
On the other hand,
in the limit of large distances
$L\,\lower3pt\hbox{$\buildrel > \over\sim$}\,L_c$,
where $\alpha_s(L^{-2})\,\lower3pt\hbox{$\buildrel > \over\sim$}\,1$,
hadronic physics is known to be described well
by an effective Lagrangian ${\cal L}[\chi,U,U^\dagger]$
which is written in terms of
the collective fields $\chi$, $U,U^\dagger$ \cite{CEO},
whose interactions are
constrained by the scale and chiral properties of the fundamental QCD
Lagrangian \cite{ellis70}.
By restricting use of the perturbative regime to
physics up to distances $L\,\lower3pt\hbox{$\buildrel < \over\sim$}\,L_c$,
and the non-perturbative effective theory to
$L > L_c$, we can
combine both regimes without double-counting, and obtain
an field theory description covering
the full range $0 < L < \infty$:
\begin{eqnarray}
{\cal L}_L[A^\mu,\psi,\chi,U]
&=& {\cal L}_L[A^\mu,\psi,\overline{\psi}]
\;+\;{\cal L}[\chi,U,U^\dagger]
\nonumber \\
&=&
-\frac{\kappa_L}{4}\,\,F_{\mu\nu, a} F^{\mu\nu}_a
\;+\;  \overline{\psi}_i \left[\frac{}{}\,\left(\frac{}{}i \gamma_\mu \partial
 ^\mu
- \mu_L\right) \delta_{ij}
- g_s  \gamma_\mu A^\mu_a T_a^{ij} \right]\, \psi_j
\nonumber \\
& & +\;
\frac{1}{2}\,(\partial_\mu \chi) ( \partial^\mu \chi )
\;+\; \frac{1}{4}\,
Tr\left[\frac{}{}(\partial_\mu U)
( \partial^\mu U^\dagger )
\right]
\;-\;V(\chi,U)
\;,
\label{Lagrangian}
\end{eqnarray}
where
$F_a^{\mu\nu}= \partial^\mu A_a^\nu -\partial^\nu A_a^\mu +
g_s f_{abc} A^\mu_b A^\nu_c$,
and summation over the color indices $a$, $i,j$ is understood.
The functions $\kappa_L$ and $\mu_L$
introduce an explicit scale($L$)-dependence in ${\cal L}_L$,
which modifies the quark and gluon properties
when $L$ increases towards $L_c$ and beyond.
In the limit $L\rightarrow 0$,
$\kappa_L=1$ and $\mu_L=0$ (neglecting the quark current masses),
and the fundamental QCD Lagrangian is recovered.
However, at larger $L$,  the bare quark and gluon fields become
dressed by non-perturbative dynamics,
and we expect $\kappa_L < 1$ and $\mu_L\ne 0$
\footnote{In fact, the short-range behaviour of
$\kappa_L$ (and similarly of $\mu_L$)
can be calculated perturbatively:
$\kappa_L = [1+ g_s^2/(8\pi)^2 (11-2 n_f/3) \ln(L\,\Lambda)]^{-1}$.
Hence both $\kappa_L$ and the dynamical quark masses $\mu_L$ vary
(as does the coupling constant $g_s$) with the renormalization
scale $\Lambda$, which we expect to be inversely
related to the confinement scale $L_c$.}.
The effective potential
$V(\chi,U)$ governs the dynamical interpolation between
short- and long-distance domains: at very small $L$, where
the typical distance between color charges is $\ll L_c$, it
is equal to the usual QCD vacuum pressure,
but as $L$ becomes large due to the
expanding spatial separation
of color charges, it simulates long-range confinement forces.
Explicit expressions for $\kappa_L$, $\mu_L$ and $V$
can be found in Ref. \cite{ms37}.
\medskip

{\bf 2.1.2 Kinetic Equations for Space-Time Evolution}
\smallskip

In Ref. \cite{ms39} it was shown how,
under certain assumptions, one can derive from the
exact quantum field theory an approximate kinetic theory.
Applied to the present case, we formulated  in \cite{ms37}
a transport-theory framework that incorporates
both partons and hadrons, yielding a fully dynamical description of
QCD matter in real time and complete phase-space.
Starting from the field equations of motion
that follow from (\ref{Lagrangian}), we obtained
the corresponding Dyson-Schwinger equations for the 2-point Green
functions
of the fields $\psi$, $A^\mu$, $\chi$, and $U$,  which
measure the time-ordered correlations
between the fields at different space-time points.
The transition from the quantum-field formulation
to a kinetic description
was then achieved by relating the Green functions
of the elementary partonic fields ($A^\mu,\psi,\overline{\psi}$)
and of the
collective hadronic degrees of freedom ($\chi,U,U^\dagger$)
to corresponding
particle densities $F_\alpha$ for the species
$\alpha = p, c, h$
of partons, clusters (prehadronic color-singlets), and hadrons,
respectively:
\begin{equation}
F_\alpha(r,k)\;\,\equiv\; \, F_\alpha (t, \vec r; \vec k, k^2)
\;\,=\;\,
\frac{dN_\alpha (t)}{d^3r d^4k}
\;,
\label{F}
\end{equation}
which are the quantum-mechanical analogues to the
classical phase-space distributions that measure the number of particles
at time $t$ in a  phase-space element $d^3rd^4k$.
The $F_\alpha$ contain the essential microscopic
information required for a statistical description
of the time-evolution of a many-particle system in
complete phase space, and provide the basis for calculating
macroscopic observables
in the framework of relativistic kinetic theory.
In terms of the particle densities (\ref{F}),
the kinetic equations derived from the Dyson-Schwinger equations
yield a set of coupled transport equations of the generic form
\begin{equation}
k_\mu \,\frac{\partial}{\partial r^\mu} \; F_\alpha (r,k)
\;=\;
\sum_{processes\;j}\;
\left[\,\hat{{\cal I}}_j^{(+)}(F_\beta)\;
-\;\hat{{\cal I}}_j^{(-)}(F_\beta)\right]
\label{Feq}
\;.
\end{equation}
These kinetic equations reflect a probabilistic
interpretation of QCD evolution in terms of successive
interaction processes $j$,
in which the rate of change of the particle distributions $F_\alpha$
in a phase-space element $d^3rd^4k$
is governed by the balance of gain (+) and loss ($-$) terms.
On the left-hand side, the covariant operator
$k^\mu \partial/\partial r^\mu = k^0\partial/\partial t -
\vec{k}\cdot \partial/\partial \vec{r}$
acting on $F_\alpha$
describes free propagation of a
quantum of species $\alpha$, whereas
on the right-hand side the interaction kernels $\hat{{\cal I}}^{(\pm)}$
are integral operators that incorporate the effects of
the particles' self and mutual interactions,
and depend functionally on the different particle densities $F_\beta$.
In general, the interactions include real and virtual emission,
absorption, scattering and coalescence. However, in the present case of
$e^+e^-$ collisions, where scattering processes and medium effects
are absent, the equations simplify considerably, and
the kernels $\hat{{\cal I}}$ fall into three categories (c.f. Fig. 1):
\begin{description}
\item[(i)]
parton multiplication through real emission processes
on the perturbative level,
$q \rightarrow q+g$,
$g \rightarrow q+\bar{q}$,
$g \rightarrow g+g$;
\item[(ii)]
parton-cluster formation through 2-parton recombinations
\footnote{This class of processes will be enlarged when
color is introduced into our approach in Sec. 2.2 below.},
$q+\bar{q} \rightarrow c_1+c_2$,
$g+q \rightarrow c+q'$,
$g+g \rightarrow c_1+c_2$;
\item[(iii)]
hadron formation through decays of the cluster excitations
into final-state hadrons,
$c \rightarrow h$,
$c \rightarrow h_1+h_2$.
\end{description}
\noindent
Whereas the branching processes (i) are the fundamental
QCD vertices, the cluster formation possibilities (ii)
depend on the specific forms of the
functions $\kappa_L$ and $\mu_L$ in the
model Lagrangian (\ref{Lagrangian}),
and the cluster decay processes (iii) are modelled
on the basis of kinematics and phase-space considerations.

In compact symbolic notation,
all interaction kernels in (\ref{Feq}) can be expressed as
convolutions of the densities of radiating or interacting particles
$F_\beta$
with the specific cross sections $\hat{I}_j$ for the processes $j$,
i.e. ${\cal I}_j = \prod_\beta F_\beta \circ \hat{I}_j$,
and one
arrives at the following closed set of balance equations for the
densities of partons $F_{p}$, clusters $F_c$ and
hadrons $F_h$:
\begin{eqnarray}
k_\mu \,\partial_r^\mu \; F_{p}
&=&
F_{p'}\circ \hat{I}(p'\rightarrow p p'')\;-\;
F_p \circ\hat{I}(p\rightarrow p' p'')  \;-\;
F_p\,F_{p'}\circ \hat{I}(p p'\rightarrow c)
\label{e1}
\\
k_\mu \,\partial_r^\mu \; F_{c}
&=&
F_p\,F_{p'}\circ \hat{I}(p p'\rightarrow c)
\;-\;
F_c\circ \hat{I}(c\rightarrow h)
\label{e2}
\\
k_\mu \,\partial_r^\mu \; F_{h}
&=&
F_c \circ\hat{I}(c\rightarrow h)
\label{e3}
\;.
\end{eqnarray}
where $k_\mu\partial_r^\mu \equiv k^\mu \partial/\partial r^\mu$,
and each of the terms on the right-hand side corresponds to
one of the above classes (i)-(iii) of processes, and is
proportional to the relevant flux,
i.e. the density of particles entering a particular vertex.
Each kernel ${\cal I}$ includes a sum over contributing
subprocesses, and a phase-space integration
weighted with the associated subprocess probability
distribution of the squared amplitude.
We discuss the physical significance of the different
kernels in more detail in Sec. 3, and
the full forms of eqs. (\ref{e1})-(\ref{e3}) and the
explicit expressions for the kernels are given in Ref. \cite{ms37}.
\medskip

{\bf 2.1.3 Calculational Scheme}
\smallskip

The set of evolution equations (\ref{Feq}) can be solved by
real-time simulation in full phase space using the
computational methods of Refs. \cite{msrep,ms3}.
In Sec. 3 we explain in more detail the
physical significance of the calculational scheme
for the specific cases of jet evolution in the processes
(\ref{ee1}) and (\ref{ee2}).

The concept of the simulation can be summarized as follows:
starting from a given initial state
\footnote{For example,
in the case of $e^+e^-$ annihilation via a virtual photon,
the initial parton state
consists of a
jet-initiating $q\bar q$ pair with invariant mass $Q$ and
flavor $f$ determined by the probability
$w_f=e_f^2/n_f(Q^2)\sqrt{1-4 m_f^2/Q^2}\,
\theta\left(Q^2 - 4 m_f^2\right)$, which
accounts for the electromagnetic charge and mass threshold.}
at time $t=0$,
the ensemble of particles is evolved in small time steps,
$\Delta t =O(10^{-3}\;fm)$, in coarse-grained
7-dimensional phase-space with
cells $\Delta \Omega = \Delta^3 r \Delta^3 k \Delta k^0$.
The time discretization is chosen to give an optimum resolution of the
particle dynamics in space and energy-momentum.
The partons propagate along classical trajectories until they interact,
i.e., decay (branching process) or recombine (cluster formation).
Similarly, the clusters so formed
travel along classical paths until they decay into hadrons.
The corresponding probabilities and time scales of interactions are
sampled stochastically from the relevant probability distributions
in the kernels $\hat{I}$ of eq. (\ref{e1})-(\ref{e3}).

With this concept, one can trace the space-time evolution of the
multi-particle system self-consistently:
at each time step, any  off-shell parton is allowed
to decay into daughter partons, with a probability determined
by its virtuality and life time
(we discuss this in more detail in Sec. 3).
Also in each step,
every parton and its nearest
spatial neighbor are considered as defining a fictious space-time bubble
in the vacuum, representing a potential candidate for
a 2-parton cluster. The probability for parton-cluster
conversion is determined by the bubble action for the
Lorentz-invariant distance $L$ between the partons.
If the two partons do convert into a cluster within any given
time slice, they disappear from that
phase-space cell, and the composite cluster appears at their
centre-of-mass point, from which it propagates on.
Otherwise the partons continue in their shower development
until the next time slice.
The final decay into hadrons of each cluster formed is simulated
analogously, and is
determined by kinematics and the available phase space.
This cascade evolution is followed until all partons
have converted, and all clusters have decayed into final-state hadrons.

In this dynamically-evolving system,
the crucial quantity that governs the  conversion
of partons into clusters and thus the structure of the final hadron state,
is the space-time scale $L(r_i,r_j)$, defined by (\ref{Ldef}),
which is the separation between two partons at $r_i$ and $r_j$,
defined by the Lorentz-invariant distance measure
\begin{equation}
\Delta_{ij} \;=\; \sqrt{ r_{ij}^\mu \; r_{ij, \mu} }
\;\;,\;\;\;\;\;
r_{ij}\; = \;r_i\, -\, r_j
\;,
\label{Dij}
\end{equation}
We identify $L$ with
the the distance $L_{ij}$ between parton $i$
and its nearest neighbor $j$:
\begin{equation}
L(r_i,r_j)\;=\; L_{ij} \;\equiv \;
\mbox{min} (\Delta_{i 1}, \ldots , \Delta_{i j}, \ldots , \Delta_{i n})
\;.
\label{L}
\end{equation}
If the separation bewteen any given parton and its nearest neighbor
approaches the confinement scale $L_c$,
cluster formation becomes increasingly likely, and
eventually occurs (c.f. Fig. 2).
The dynamics of this cluster formation is described by
the kernels $F_i F_j \circ \hat{I}(i j\rightarrow c)$
in eqs. (\ref{e1}) and (\ref{e2}), that determine the
probability for the coalescence of two partons $i$, $j$ to form
clusters $c$. It is modelled by a distribution of the form
\begin{equation}
\Pi_{ij\rightarrow c}\;= \;{\cal C}(L_{ij})\; \left(\frac{}{}
1\,-\, \exp\left(-\Delta F\;L_{ij}\right)
\right)
\;,
\label{Pi2}
\end{equation}
where ${\cal C}$ modifies the small-$L_{ij}$ behaviour where the
exponential form is not appropriate \cite{ms37},
$L_{ij}$ is the 2-parton separation defined by (\ref{L}),
and $\Delta F$ is the local change in the free energy
of the system that is associated with the
conversion of the partons to clusters.
The latter is determined by the specific form of the confinement
potential $V$ in the model Lagrangian (\ref{Lagrangian}).
Our previous analysis \cite{ms37} indicates that
we can parametrize the resulting probability density (\ref{Pi2}) as
\begin{equation}
\Pi_{p_ip_j\rightarrow c}\;=\;
\left\{
\begin{array}{cl}
0
&\;\;\;\;\;\mbox{if $L_{ij} \;\le \;L_0$} \\
1\;-\;\exp\left(\frac{L_0-L_{ij}}{L_c-L_{ij}} \right)
&\;\;\;\;\;\mbox{if $L_0 \;< \;L_{ij}\;\le\;L_c$} \\
1
&\;\;\;\;\;\mbox{if $L_{ij}\;>\;L_c$}
\end{array}
\right.
\;,
\label{Pi3}
\end{equation}
where $L_c = 0.8\; fm$ is the value for the
confinement length scale that fits best the Bose-Einstein
correlation data \cite{ms37}, and $L_0 = 0.6\;fm$.
The transition interval
$[L_0,L_c]$ arises from the finite, thin-walled potential barrier
of the model potential $V$ in (\ref{Lagrangian}),
that separates the perturbative and hadronic vacua.
Its effect is to yield a small, but non-vanishing,
probability (\ref{Pi3}) for partons
to convert to clusters while their separation is still smaller,
but close to, the confinement length scale $L_c$, as
illustrated in Fig. 2.
This scheme separates perturbative parton evolution and
non-perturbative confinement dynamics in a self-regulating manner,
because in the mean it bounds the partons' virtualities
from below:
\begin{equation}
k^2 \;\ge \; \mu_0^2 \;=\; O(L_0^{-2})
\label{mu0}
\end{equation}
so that the condition
$\alpha_s(k^2) \le \alpha_s(\mu_0^2) \ll 1$
is imposed dynamically
\footnote{
Although this condition determines  the mean $\mu_0$ to come out
around 1 GeV, statistical fluctuations occur with
the distance $L_{ij}$ of two partons $i$ and $j$ in the interval
$L_c < L_{ij}^{-1} < \mu_0$, in which case we cut off the parton radiation
below $\mu_0$, and let the partons propagate freely without branching,
until they eventually coalesce.
}
that clearly defines the notion of a perturbative, dressed  parton, as opposed
to
a non-perturbative clustered parton.

The essential parameter in this description is
the confinement scale $L_c$, defined by (\ref{Lc}).
This is in principle related to the critical temperature
in the finite-temperature QCD phase transition, but is in
practice subject to some uncertainty.
In Ref. \cite{ms37} we tested different choices
by studying the evolution of hadronic $e^+e^-$ events,
and found that the details of the transition are rather insensitive
to reasonable variations in $L_c$.
However, the value $L_c = 0.8$ $fm$
gives the most favorable description of the Bose-Einstein
correlations observed on the $Z^0$ peak, as well as other
experimental data.

This completes the summary of Ref. \cite{ms37}. We now turn to the
extension of our model to take better
account of color degrees of freedom.
\medskip

\noindent {\bf 2.2 Color Degrees of Freedom and Correlations}
\smallskip

In our previous description of parton evolution and hadron formation,
we did not take explicit account of the color degree of freedom,
which is after all the origin of the confinement mechanism.
Our Ansatz for confinement picture was based exclusively
on the dynamically-evolving space-time separations (\ref{L}) of
nearest-neighbor color charges in the parton cascade, rather than on
the details of the color
structure of the ensemble of produced gluons, quarks and antiquarks.
Our underlying justification for this reasoning
was the pre-confinement property \cite{preconf}
of jet evolution in perturbative QCD, according to which
the partons emitted during a cascade tend to arrange themselves
into a collection of color-singlet
systems, which may be viewed as minimal pre-hadronic units.
The following two important features characterize
the pre-confinement phenomenon \cite{bassetto}.
First, the color-singlet subsytems of partons so formed, typically
have masses of the order of 1 - 2 GeV, independent of
the total energy of the system as a whole. Second, in the
leading-logarithmic approximation,
the color quantum numbers are naturally ordered in such a
way that the partons which form a color singlet are topologically
nearby in a cascade (c.f. Fig. 3), and have a finite
space-time separation in the color-singlet rest-frame.

It is therefore suggestive to identify these color-singlet
systems with the prehadronic clusters in our approach,
as we did implicitly in our previous work \cite{ms37}.
It should be emphasized, however, that
this correspondence between the color
and space-time structures of a parton cascade is not an
equivalence, but holds only in the average \cite{LPHD,caneschi,marchesi,ms18}:
whereas the color structure of a cascade tree provides in principle
exact microscopic information about the flow of color charges,
the space-time structure is based in our model on the statistical kinetic
description of parton emission and the nearest-neighbor search,
which may be subject to fluctuations that deviate from the
exact color flow.
This issue becomes increasingly important when more particles populate
a phase-space region (possibly already for the small-$x$ region in
deep-inelastic scattering, and certainly for hadron-nucleus
or heavy-ion collisions).
Intuitively, we would expect it then to become increasingly
likely that nearest neighbors in momentum space would not necessarily form
a color-singlet. It could also be that the `natural' color-singlet
partner for a given parton within the same cascade (its `endogamous'
partner) might actually be dynamically-disfavoured by comparison with
a color-singlet partner from a different, but overlapping, cascade
(an `exogamous' partner). A crucial first step in addressing the
latter possibility has been the scenario for `color reconnection',
first explored phenomenologically in \cite{GPZ}.
We believe that our model, which is capable of following
simultaneously the space-time developments of overlapping cascades,
provides a potentially-interesting alternative tool for addressing
these issues. Moreover, our nearest-neighbor criterion
(\ref{L}) provides a plausible way of using this tool to
resolve questions of `color reconnection'.

Accordingly we now incorporate color flow into our model, within
the general framework of the probabilistic parton cascade
description of QCD jets that works
so impressively well at high energies (see e.g. \cite{webber95}).
We assign to each parton a color ($C$) and an anticolor ($A$) label:
\begin{equation}
(\,C,\,A\,) \;\,=\;\,
\left\{
\begin{array}{cl}
(i, 0) & \;\;\;\;\;\; \mbox{quarks} \\
(0, j) & \;\;\;\;\;\; \mbox{antiquarks} \\
(k, l) & \;\;\;\;\;\; \mbox{gluons}
\end {array}
\right.
\;,\;\;\;\;\;\;
(i,j,k,l \;=\; 1, \ldots , N_c)
\;.
\label{color}
\end{equation}
It is straightforward to specify the
color flow at each elementary vertex
for parton branching, and in parton recombinations to form
clusters, as shown in Fig. 4 and Fig. 5, respectively.
Since it is the nature of a probabilistic description that it only
depends on the local properties of the system
(i.e. non-local interference
effects are assumed to be negligible), this scheme
provides an unambigous local tracer of the color flow,
in complete analogy to energy-momentum and flavor flow.
It is a defect of this approach that there are
nine, rather than eight, gluon color
states. We have made numerical studies varying $N_c$, and found
no significant effects in our results.
However, this point remains a
conceptual issue that needs to be thought through
more carefully. In particular, it would be desirable to understand
the relation of our approach
to the $1/N_c$ expansion and the dominance of planar diagrams
in high-energy QCD \cite{bassetto}.

To see how important color correlations can be, we
distinguish the following
three scenarios, which differ in their level of accounting
for the color flow in conjuction with the space-time
evolution.
\begin{description}
\item[(I)]
{\it Color-Blind} Scenario:
The color degrees of freedom (\ref{color}) are
ignored. Cluster formation of
partons is based solely on the
nearest-neighbor criterion (\ref{L}).
It is assumed that pairs of nearest-neighbor partons are
on the average also color-singlets, and therefore can
form one, or two pre-hadronic clusters.
\item[(II)]
{\it Color-Singlet} Scenario:
As (I), but with the additional restriction that the
color degrees of freedom (\ref{color}) of two
partons that are potential candidates to form a cluster
must add up to a color-singlet combination
\footnote{
Color-neutral systems need not be color singlets:
for example, the color-octet representation of $SU(3)_c$
contains two color-neutral members, meaning that they  possess zero
eigenvalues of the non-comuting generators $\tau_3$ and $\tau_8$.
In the classical approximation of parton showers,
the color phase of the amplitude for each emission is random,
so that the final-state partons are
statistically distributed among the possible $SU(3)_c$
representations. Therefore, in this naive approximation,
e.g., a color-neutral $q \bar q$ pair
has equal probabilities for being a color singlet or a color octet.
}.
This corresponds to considering only the $2\rightarrow 2$
`color-singlet'
processes of the left column in Fig. 5.
\item[(III)]
{\it Color-Full} Scenario:
No restrictions whatsoever are imposed
on the colors of pairs of partons participating
in the formation of a cluster,
meaning that, in contrast to (II), all the
processes illustrated in Fig. 5 are
included, i.e.
$2\rightarrow 2$, $2\rightarrow 3$, and $2\rightarrow 4$.
If the space-time separation (\ref{L}) of two nearest-neighbor partons
allows  coalescence, they can
always produce one or two color-singlet clusters, accompanied, if
necessary,  by the emission
of a gluon or quark that carries away any unbalanced net color.
\end{description}
We note that the elementary branching processes
remain the same as before except for the additional
color information, but that the cluster formation
dynamics depends on the scenario considered:
scenario (I) corresponds to our original approach \cite{ms37} as
reviewed earlier in this Section,
scenario (II) is the most restrictive one, in that it allows
only a minimal number of processes, whereas scenario (III) is
the most diverse, in the sense that it increases the original
number of processes by allowing additional parton emission.

In the following Section
we compare the application of the three scenarios
to $e^+e^-$ events of the types (\ref{ee1}) and (\ref{ee2}),
i.e. via $Z^0 \rightarrow 2 \;jets$
and $W^+W^- \rightarrow 4 \;jets$.
As we will show, our results for the overall
space-time development, as well as for global
event measures such as multiplicities or momentum spectra, are
rather insensitive to the differences between
the above color structure scenarios.
However, as we shall see,
more sensitive observables may well distinguish between them,
in particular the measurement of the
$W$ mass in 4-jet events of the type (\ref{ee2}).
\bigskip
\bigskip

\noindent {\bf 3. APPLICATION TO HADRONIC {\boldmath $e^+e^-$} EVENTS
AT LEP}
\bigskip

This Section is devoted to the Monte Carlo simulation of the
hadronic $e^+e^-$ event types (\ref{ee1}) and (\ref{ee2})
within the variants of our model outlined in Sec. 2.1.3.
Between 50000 and 150000 events were accumulated in each simulation,
depending on the statistical accuracy required for each calculated
observable. Each event generated
was traced in time steps $\Delta t = 10^{-3}$ $fm$ from
the decay of the jet-initiating particle
($\gamma^\ast$ or $Z^0$)
at $t_0=0$, up to a final time $t_f =20$ $fm$ afterwards,
when the parton showers have
completed their conversion into final-state hadrons.

We study first the case of 2-jet evolution
in  $e^+e^- \rightarrow  Z^0 \rightarrow hadrons$
on the $Z^0$ peak at the nominal energy of LEP1, $\sqrt{s}= 91$ GeV, and
subsequently the more complex case of 4-jet evolution in
$e^+e^- \rightarrow  W^+W^- \rightarrow hadrons$
at the nominal LEP2 energy of $\sqrt{s} = 170$ GeV.
We focus particular attention on the following two issues:
a) the characteristics
of the space-time evolution, which we discuss in the
light of the `inside-outside cascade' picture \cite{bj,marchesi},
and b) the impact of color correlations
on observable quantities in experiments,
comparing the scenarios (I)-(III) defined at the end of
 the previous Section.
\bigskip

\noindent {\bf 3.1 {\boldmath
$e^+e^-  \rightarrow  Z^0 \rightarrow q_1 \bar{q}_2 \rightarrow hadrons$}
at {\boldmath $\sqrt{s} = 91$} GeV}
\medskip

Let us return to the kinetic equations (\ref{e1})-(\ref{e3}) that
describe for $e^+e^-$ collisions the cascade evolution of
the mixed particle system
consisting of gluons, quarks, antiquarks, prehadronic clusters and
hadronic states.
We discuss the physical significance of the equations and the
emerging space-time picture first
for the simpler event type (\ref{ee1}) of an evolving 2-jet configuration
produced by a virtual $Z^0$.
The additional features of the event type (\ref{ee2}) are addressed
in Sec. 3.2.
\medskip

{\bf 3.1.1 General Discussion}
\smallskip

Consider the production of the initial quark-antiquark pair
$q_1\bar{q}_2$
with opposite flavor and color quantum numbers
on the $Z^0$ peak in the restframe
of the decaying $Z^0$ at an
invariant mass $Q= \sqrt{s} = E_{cm}$.
Let us for the moment ignore the parton-cluster hadronization mechanism,
and focus on the early stage of parton cascade evolution.
In this case only eq. (\ref{e1}) is relevant, and the
right-hand side includes solely the space-time version of the
leading-log QCD-evolution kernels \cite{ms37}
(the third term would be absent).
That implies, initially,
free streaming of the initial $q_1$ and $\bar{q}_2$
which recede  back-to-back along straight-line trajectories
away from the $Z^0 \rightarrow q\bar{q}$ vertex,
which we choose as
$r_0^{Z^0} := (t_0, \vec{r}_0) = (0,\vec{0})$.
Their velocities
are $\vec{\beta}_{1,2} = \pm \vec{p}_{cm}/E_{cm}$, where
$\vec{p}_{cm} = \vec{k}_1 = - \vec{k}_2$,
$E_1 = E_2 = \sqrt{p_{cm}^2 + m_q^2}$.
The $q_1$ and $\bar{q}_2$ are off-shell with a time-like virtuality
$k_i^2 < k_{i\;max}^2 = Q^2/4$ ($i = 1,2$), and hence live only
for a finite time before they decay by gluon emission.
In the $Z^0$ rest-frame, the characteristic
life time is \cite{marchesi,ms18}
\begin{equation}
t_p(x_i,k_i^2) \;\,=\;\,\;\gamma_i \, \tau_i
\;\simeq\; E_i/k_i^2 \;=\; \frac{x_i Q}{2\,k_i^2}
\;,
\label{ti}
\end{equation}
where the
$\gamma_i$ are the Lorentz factors of $q_1$ and $\bar{q}_2$,
the $\tau_i \simeq 1/\sqrt{k_i^2}$ their proper life times as given
by the uncertainty principle, and the $x_i = E_i/E_{cm} = 2E_i/Q$
their energy fractions.
When and where in space time the decays of the
$q_i \equiv q_1, \bar{q}_2$
occur is obtained statistically from the kernels
$F_{q_i}\circ\hat{I}(q_i\rightarrow q_i' g_i'')$ in eq. (\ref{e1}),
that incorporate the probability distributions
for the branchings of momenta $k_i \rightarrow k_i' + k_i''$
with energy ratios $z=x_i'/x_i$ and $1-z = x_i''/x_i$,
\begin{equation}
\Pi_{q_i\rightarrow q_i'g_i''} \;=\;
\frac{ \alpha_s\left[(1-z)k_i^2\right]}{2 \pi}
\;\,\gamma_{q_i\rightarrow q_i'g_i''}(z)
\,\; T_{q_i}(x_i,k_i^2)\;
\;,
\label{pi1}
\end{equation}
where the argument of $\alpha_s$ is set by the momentum scale
$(1-z) k_i^2 \approx k_\perp^2$ associated
with the vertex \cite{bassetto}, the function $\gamma(z)$ denotes the
usual DGLAP energy- ($z$-) distribution \cite{DGLAP,dok80},
and the life-time factor $T_{q_i}$ expresses the probability
for the quark or antiquark  $q_i$ of virtuality $k_i^2$ to decay within
a time interval $\Delta t$ in the $Z^0$ rest-frame,
\begin{equation}
T_{q_i}(x_i,k_i^2)\;=\; 1\;-\; \exp\left( - \frac{\Delta
 t}{t_{q_i}(x_i,k_i^2)}\right)
\;,
\label{ltf}
\end{equation}
where $t_{q_i}(x_i,k_i^2)$ is given by (\ref{ti}).

The daughter partons of the initial branchings
$q_1 \rightarrow q_1' g_1''$ and $\bar{q}_2
\rightarrow \bar{q}_2' g_2''$ then
follow the same pattern, with $k_i'^{\,2}, k_i''^{\,2} < k_i^2$
and corresponding life times $t(x_i',k_i'^{\,2})$
and $t(x_i'',k_i''^{\,2})$.
The elementary branchings that can occur are
$q\rightarrow qg$, $\bar{q} \rightarrow \bar{q}g$, $g\rightarrow gg$,
$g\rightarrow q\bar{q}$. We employ the (modified)
leading-logarithmic
description \cite{MLLA,dok80} in conjunction with the
angular ordering \cite{bassetto} of
subsequent emissions, which is equivalent to an ordering in virtualities
at large $x$, but accounts for interference effects among soft gluons at
small $x$.

The parton shower thus develops as illustrated in Fig. 6, where the
time estimate for the $n$'th branching is
\begin{equation}
t_n\;=\; \sum_{i=1}^{n} t(x_i,k_i^2)\;=\; \frac{Q}{2}
\;\sum_{i=1}^n \frac{x_i}{k_i^2}
\label{tn}
\;.
\end{equation}
In the leading-logarithmic approximation and in the limit of small $x$,
these considerations lead to the following
analytical result for the average time development of the
parton cascade \cite{marchesi}:
\begin{equation}
\langle \,t_p(x,k^2) \rangle \;
\sim \;
a\;\frac{x Q}{2 k^2} \;\exp\left( -\,b\,
\sqrt{\ln\left(\frac{1}{x}\right)}\right)
\label{tav}
\;,
\end{equation}
in the limit $x\ll 1$ and $k^2\simeq\mu_0^2$,
where $\mu_0 \simeq 1$ GeV, and $a$ and $b$ are slowly-varying functions
of $Q^2$ and $k^2$. Hence, as $x\rightarrow 0$,
the average time for parton production
$\langle t_p\rangle \rightarrow 0$,
implying that soft (small-$x$) partons are emitted earlier than
the fast (large-$x$) ones.
This fact is a consequence of time dilation, and is known as the
`inside-outside cascade' \cite{bj,kogut73}.
We will return to this analytical property when discussing
our numerical results below.

Because the virtualities are strongly
ordered in the leading-logarithmic approximation,
and decrease in a time-like cascade,
the quanta become increasingly `dressed':
$k^2 \rightarrow \Lambda_{QCD}^2$
as time progresses: $t\rightarrow \infty$. In the absence
of confinement, they would eventually reach mass shell.
However, the confinement length scale $L_c$, defined by (\ref{Lc}),
limits the time that the partons can evolve perturbatively, so that
they never manage to reach mass shell, but first
undergo the non-perturbative hadronization mechanism.
It is here that the cluster formation scheme outlined
in the previous section terminates
locally the perturbative evolution,
by coalescing two nearest-neighbor partons $i$ and $j$,
if their separation $L_{ij}$ (\ref{L}) in their
common rest-frame approaches or exceeds $L_c$
(as described before in Sec. 2.1),
thereby simulating the screening of the
color charges \cite{caneschi,marchesi}, which is responsible for
confinement.

Once two partons $i$ and $j$ coalesce,
the combined system with invariant mass
$m_{ij} \equiv \sqrt{(k_{i}+k_{j})^2}$ is decayed
into the outgoing cluster(s) and possible additional partons
by using 2- or 3-body decay kinematics and phase-space,
depending which of the actual processes of Fig. 5 occurs.
In the case of the 2-cluster processes $i+j\rightarrow c_1+c_2$,
consistent with the symmetry under exchange of $c_1$ and $c_2$,
the mass $m_{c}$ of the first cluster $c_1$ is sampled from an exponential
distribution $\propto \exp[-m_c/(m_{ij}-m_c)]$, and the second
cluster $c_2$ carries off the remaining energy-momentum according
to 2-body decay kinematics.
An analogous
procedure is used for processes with the emission of one parton $l$,
$i+j\rightarrow c+l$ by replacing $c_2$ with $l$.
In the case of  a cluster $c$ accompanied by two emitted partons $l_1$, $l_2$,
$i+j\rightarrow c+l_1+l_2$, again the cluster mass is generated,
and then  3-body decay kinematics determines the
four-momenta of the two outgoing partons in the rest-frame of $i+j$.
In each case, the outgoing particles are then boosted from the
$i+j$ rest-frame back to the $Z^0$ frame.
Whereas any outgoing parton participates in the continuing
perturbative shower development, each produced cluster
with determined four-momentum $k^\mu_c=(E_c,\vec{k}_c)$, mass
$m_c = k_c^2$ and proper life time $\tau_c = 1/m_c$, propagates along
a straight path with velocity $\vec{k}_c/E_c$ until
it converts into final-state hadrons after a life time
\begin{equation}
t_c(E_c,m_c)\;=\; \gamma_c \;\tau_c
\;\simeq\;
\frac{E_c}{m_c^2}
\label{t2}
\;,
\end{equation}
This cluster-decay scheme is embedded in the
kernels $F_c\circ\hat{I}(c\rightarrow h)$
of eqs. (\ref{e2}) and (\ref{e3}),
and is modelled \cite{ms37} along the lines originally proposed
in \cite{webber84}:
If a cluster is too light to
decay into a pair of hadrons, it is taken to represent
the lightest single meson that corresponds to its
partonic constituents. Otherwise, the cluster
decays isotropically in its rest-frame into
a pair of hadrons, either mesons or baryons, whose combined
quantum numbers correspond to its partonic constituents.
The corresponding  decay probability is chosen to be
\begin{equation}
\Pi_{c\rightarrow h}\;=\;
\;T_c(E_c,m_c^2) \;
\;\,{\cal N}
\int_{m_h}^{m_c}
\frac{dm}{m^3}\;\exp\left(-\frac{m}{m_0}\right)
\;,
\label{pi3}
\end{equation}
where ${\cal N}$ is a normalization factor,
and the integrand is a Hagedorn spectrum \cite{hagedorn} that
parametrizes quite well
the density of accessible hadronic states below $m_c$ which are
listed in the particle data tables, and $m_0 \simeq m_{\pi}$.
In analogy to (\ref{ltf}), $T_c$ is a
life-time factor giving the probability
that a cluster of mass $m_c^2$ decays  within
a time interval $\Delta t$ in the $Z^0$ rest-frame
\begin{equation}
T_c(E_c,m_c^2)\;=\; 1\;-\;
\exp\left( - \frac{\Delta t}{t_c(E_c,m_c^2)}\right)
\;,
\label{lft2}
\end{equation}
with the Lorentz-boosted life time $t_c(E_c,M_c^2)$ given by (\ref{t2}).
The only exception to the above rules is that
if a cluster is very heavy, $m_c > m_{fiss} = 4$ GeV,
it undergoes one or more fissions before the products then decay into
hadron pairs \cite{webber84, ms37}.
In this scheme, a particular cluster
decay mode is obtained from (\ref{pi3})
by summing over all possible decay channels,
weighted with the appropriate spin, flavor, and
phase-space factors, and then choosing the actual decay mode
acording to the relative probabilities of the channels.
\smallskip

In summary, we perform a
complete simulation of the collision process which
follows the microscopic space-time development from
the initial $q\bar q$ of the $Z^0$ decay, via
parton-shower evolution and parton-cluster formation and decay,
all the way to the final-state hadrons. The origins of all the
hadronic observables are therefore recorded in
the particular space-time history of the system.
\medskip

{\bf 3.1.2 Numerical Results}
\smallskip

We now discuss the results of our numerical simulations
for $e^+e^-$ collisions at 91 GeV, paying particular attention
to the comparison between the color
scenarios (I)-(III) defined in Sec. 2.2.
We chose for each of these scenarios an appropriate value for
the minimum required parton virtuality
during the cascade evolution, $\mu_0$ of (\ref{mu0}), so as
to reproduce the {\it charged} multiplicity, which is determined
quite accurately by experiment: $n_{ch} = 20.95 \pm 0.88$ \cite{lep}.
Table 1 summarizes the
values for $\mu_0$ required by this procedure, as well as
the calculated multiplicities of partons and hadrons, which are
compared with with experimental data \cite{MC} where available.
One observes that the color-blind scenario (I) requires
the smallest value of $\mu_0$,
whereas the color-full scenario (III) needs the largest value.
Correpondingly, the parton multiplicities are largest (smallest)
in these two extreme scenarios.
However, the {\it total} hadron multiplicity does not follow the same
pattern, since scenario
(III) produces the largest yield, giving the smallest
parton-to-hadron ratio. This is due to the presence of
cluster-formation processes involving additional parton emission,
which lead in the average to lower-mass clusters, because the
emitted parton(s) take away energy-momentum. Since the
total energy of the system is conserved, in the end
a larger number of clusters with lower mean masses
are been produced (c.f. Table 2),
which leads naturally to an increase of the hadron multiplicity.
Regarding the individual hadron multiplicities, there are
significant differences between scenarios (I)-(III)
in the particle composition of the final state,
but the numbers are generally within reasonable
range of the available experimental data, even in the absence of
further tuning of the scenarios.

Fig. 7 reflects the emerging time-evolution pattern of
the dynamical development of the mixed-particle system,
presenting more details of the parton showering and of
parton-hadron conversion. The four plots show the relative
proportions of partons and hadrons,
the conversion of the total energy, the fractions that the partons
and hadrons carry of the longitudinal momentum fractions
along the 2-jet axis, and the total
produced transverse momentum perpendicular to this axis.
One can conclude, first, that the evolution can
be divided crudely into three stages:
(i) a {\it parton shower stage}
($\,\lower3pt\hbox{$\buildrel < \over\sim$}\, 0.5$ $fm$),
(ii) a {\it parton-hadron conversion stage} ($\approx 0.5 - 5$ $fm$),
and, (iii) a {\it hadron stage}
($\,\lower3pt\hbox{$\buildrel > \over\sim$}\, 5$ $fm$).
Second, it is evident that at a macroscopic level
the overall space-time evolution is only marginally different for
scenarios (I)-(III), implying that the gross features
of the dynamical parton-hadron
conversion are primarily determined by kinematics
and the way in which the particles occupy phase space,
and to a much lesser extent
by the role of the internal color degrees of freedom.

Fig. 8 shows that the differences among
the color flow scenarios (I)-(III)
have a particularly small impact on the resulting
cluster size- and mass-spectra, as well as on
the momentum distributions of the hadrons
that emerge from the cluster decays.
This is also reflected in the cluster properties listed in Table 2,
which supplements the previous Table 1.
Table 2 provides, in addition, a listing of the relative contributions
of the various parton-cluster formation processes defined in
Fig. 5. Comparing the color-singlet and color-full scenarios
(II) and (III), respectively,
one may conclude that most processes in the latter case are of
non-singlet type, accompanied by color emission.
On the other hand, scenarios
(I) and (III) give very similar total rates for
$gg$, $q\bar q$ and $qg$ cluster processes, whereas
scenario (II) clearly differs. It is the most restrictive
in the sense that it allows only color-singlet clustering, and
consequently suppresses substantially direct $gg$ or $q\bar q$
recombinations, because the color non-singlet configurations
among these are vetoed. This yields an increase of the
relative $qg$ recombination
rate, since here the outgoing quark can carry off the net color.

The conclusion from these results is that
global quantities associated with
the bulk multi-particle system, averaged over many events,
do not exhibit obvious spectacular effects of the color flow structure.
In order to extract an observable effect, one has to study
particular event shapes or other measures that are sensitive to
the local color structure of the system. In the next Section
we give examples in the context of the $W$ mass determination
in events of the type (\ref{ee2}),
and find significantly distinct results.
\smallskip

Before doing so, however, we return to the general space-time structure
of the event evolution, and explore whether our approach is consistent
with the `inside-outside' cascade picture that one expects from
both intuitive considerations \cite{bj,kogut73}
and the formal arguments \cite{caneschi,marchesi}
reviewed earlier. We see in Figs. 9 and 10 that the answer is `yes'.
In Fig. 9 we show the time developments of the parton
spectra versus rapidity $y$, longitudinal direction $z$ along the
2-jet axis, transverse momentum $p_\perp$, and perpendicular
direction $r_\perp$. In Fig. 10, the corresponding time
developments of the hadron spectra are plotted.
The various curves correspond to `snapshots'
taken during the course of the evolution at the indicated
discrete time intervals in the $Z^0$ rest-frame.
The simulations clearly exhibit the
expected `inside-outside' character of the
rapidity ($y$) and longitudinal ($z$) space-time evolution. The central
$y$- and $z$-region is populated the earliest, first by the
bulk of small-$x$ partons with
\footnote{
The general relation between rapidity $y$ and
energy (or momentum) fraction $x$ is
$y = -\ln(1/x)- 1/2\, \ln((k^2+k_\perp^2)/Q^2)$, which implies
approximately $y \simeq |\ln x |$ for sufficiently large $k^2$.
}
$y\simeq \ln x$,
and then by hadrons that result
from the disappearing partons in the cluster formation process.
This property is in accord with the perturbative QCD result (\ref{tav})
calculated in the leading-logarithmic approximation.
The regions at larger $y$ and $z$ are gradually populated later,
as the system expands.
Furthermore, one sees that the time-development
in transverse momentum- and coordinate-space
proceeds by diffusion in both $p_\perp$ and $r_\perp$.
The average absolute value of $p_\perp$ per particle decreases (although
the total transverse momentum grows, c.f. Fig. 7), and so
the slope of the distribution steepens with progressing time.
This is a random walk effect, which is mirrored in
the diffusion in $r_\perp$ with a broadening particle
population prependicular to the jet axis.

Finally, Fig. 11 gives a 2-dimensional visualization of this
`inside-outside' picture in $log(t)$ versus $y$ (top), and
$z$ (bottom). We show the particle ratios
\begin{equation}
R_{y, z} (t) \;=\;\frac{\rho_{y,z}^{(h)}}
{\rho_{y,z}^{(p)}+\rho_{y,z}^{(h)}}\;,
\label{phratio}
\end{equation}
where
\begin{equation}
\rho_{y}^{(p,h)} \;\equiv\; \frac{d N^{(p,h)}(t)}{dy}
\;\;,\;\;\;\;\;\;\;\;\;\;
\rho_{z}^{(p,h)} \;\equiv\; \frac{d N^{(p,h)}(t)}{dz}
\;,
\end{equation}
i.e., the number of hadrons divided by the total number of particles,
which measures the rate of hadron yield locally in
$y$ and $z$. Initially zero, the ratios (\ref{phratio})
must approach unity as time progresses.
The plots again confirm that time dilation from the
particles' local, comoving frame to the $Z^0$ rest-frame
yields a picture in which hadrons are first produced
at small $y$ and $z$ where most partons are emitted at early time,
and then `eat' their way out to larger rapidity and spatial
distance from the initial $Z^0$ vertex.
\bigskip

\noindent {\bf 3.2. {\boldmath
$e^+e^-  \rightarrow \gamma^\ast / Z^0 \rightarrow W^+W^-\rightarrow
q_1 \bar{q}_2 q_3 \bar{q}_4\rightarrow hadrons$}
at {\boldmath $\sqrt{s} = 170$} GeV}
\medskip

The 4-jet evolution initiated by the
$q\bar q$ decays of a $W^+$ and a $W^-$ that were produced by
an intermediate $\gamma^\ast$ or $Z^0$ embodies new features
compared to the previous case of 2-jet evolution
of a single $q\bar q$-pair.
In particular, since the $W^+$ and $W^-$ decay vertices overlap in space,
the initial state is a single QCD ``hot spot'' in which
collective non-perturbative hadronization effects can be important.
We proceed as in the preceding Section, by discussing first the general
space-time picture of the process, proceeding later to investigate the
different color correlation effects in scenarios (I)-(III),
with an eye to the
experimental measurement of the $W$ mass through jet reconstruction.
\medskip

{\bf 3.2.1 General Discussion}
\smallskip

Again we describe the dynamics in the overall center-of-mass frame,
that is the rest-frame of the $\gamma^\ast/Z^0$
with invariant mass $Q = E_{cm} = \sqrt{s}$.
The produced $W^+$ and $W^-$ are generally off mass shell,
meaning that their masses $m^\pm$ are unequal, and
differ from the nominal on-shell value, which we take to be
$m_W = 80.22$ GeV \cite{pada}. The distribution of produced masses $m^\pm$
can be described by a relativistic Breit-Wigner distribution,
\begin{equation}
\Pi(m^\pm)\;=\;
const. \; \frac{(\Gamma_W/m_W)^2}{1-(m_W/m^\pm)^2 \;+\; (\Gamma_W/m_W)^2}
\label{breitwigner}
\;,
\end{equation}
where the full width is taken to be $\Gamma_W = 2.12$ GeV \cite{pada}.
As a consequence, the energies
$E^\pm = (s\pm ((m^+)^2 - (m^-)^2)/\sqrt{4s}$ differ from those
expected in the naive on-mass-shell case where $m^+=m^-=m_W$.
Furthermore, although the three-momenta satisfy
$\vec{p}^{\,+} = - \vec{p}^{\,-}$, the $cm$-momentum
$p_{cm}\equiv |\vec{p}^{\,+}| = |\vec{p}^{\,-}|$ deviates \cite{SK}
from the naive case: in fact,
$p_{cm} \simeq 20$ GeV at $\sqrt{s} = 2 m_W \simeq 160$ GeV,
rather than the naive value
$p_{cm}^{naive} = 0$, and
$p_{cm} \simeq 30$ GeV at $\sqrt{s} = 170$ GeV.

With this kinematic situation in mind, we describe the
evolution of the system as follows.
As before, we choose the
space-time position of the $\gamma^\ast/Z^0$-decay vertex to be
$r_0^{\gamma^\ast /Z^0} := (t_0, \vec{r}_0) = (0,\vec{0})$. The produced
$W^+$ and $W^-$ recede in opposite directions with different velocities
$\vec{\beta}^{\pm} = \pm \vec{p}_{cm}/E^\pm$, and different
Lorentz boost factors $\gamma^\pm = E^\pm/m^\pm$.
This causes variations in the life times of the
$W^+$ and $W^-$ in the center-of-mass frame:
\begin{equation}
\langle \,t^\pm\,\rangle \;=\; \gamma^\pm\;\tau^\pm
\label{tpm}
\end{equation}
and the two particles subsequently decay at space-time vertices
separated by
\begin{equation}
r^\pm \;\equiv\; (t^\pm,\vec{r}^\pm) \;=\;
\gamma^\pm \tau^\pm \,(1, \vec{\beta}^\pm)
\label{rpm}
\end{equation}
from $r_0^{Z^0}=(0,\vec{0})$.
The proper life times $\tau^\pm$ are given by a distribution
\begin{equation}
\Pi(\tau^\pm)\;=\;
\frac{1}{\tau_W(m^\pm)} \;\exp\left(-\frac{\tau^\pm}
{\tau_W(m^\pm)}\right)
\label{pitau}
\end{equation}
with
\footnote{
As has been noted in \cite{SK},
this expression, which follows from the assumed Breit-Wigner distribution
(\ref{breitwigner}), is not exact. A general discussion of
unstable particle life-times can be found in, e.g., Ref. \cite{wigner}.
}
\begin{equation}
\tau_W(m^\pm)\;=\;
\frac{m^\pm}{\sqrt{(m^{\pm\;2}-m_W^2)^2 \,+\, (\Gamma_W \,m^{\pm\;2}/m_W)^2}}
\label{taupm}
\;.
\end{equation}
Because of the short proper life time resulting from
(\ref{taupm}), and the fact that the
boost factors are not large near the $W^+W^-$ threshold at
the nominal LEP2 energy of $\sqrt{s}=170$ GeV,
the vertices are typically very close \cite{SK}:
on the average, the time separation is $|t^+-t^-| \simeq 0.08$ $fm$
and the spatial separation is
$|\vec{r}^{\,+}-\vec{r}^{\,-}| \simeq 0.05$ $fm$.
This situation is illustrated in Fig. 12, which is meant
to represent the creation of a single QCD ``hot spot'', rather than
of two independent ``hot spots''.

The $W^+$ and $W^-$ each decay into a time-like $q\bar q$-pair, and
both the $q_1\bar{q}_2$ and the $q_3\bar{q}_4$ are produced
off-shell with maximum virtualities
$k_{1,2}^2 < (m^+)^2/4$ and $k_{3,4}^2 < (m^-)^2/4$.
Assuming that perturbative interference effects
between the $W^+$ and $W^-$ decays are negligible \cite{SK},
we then visualize each $q\bar q$-pair as a
2-jet system that initiates an independent
time-like parton shower, whose evolution we
describe in coherent angular-ordered manner using
the same scheme as in the previous Section:
the two cascades develop by successive parton emission,
followed by cluster formation and final cluster decay into hadrons.
There is, however, an important physical difference as compared
to the case of a single 2-jet evolution.
Due to the close proximity of the initial $q_1\bar{q}_2$
and $q_3\bar{q}_4$ pairs, the
associated 2-jet systems overlap in space-time, which results
in overlapping phase-space populations from the two showers,
implying that parton correlations are likely to affect the
recombination of partons to color-singlet clusters,
and hence the hadron yield.
This leads to a mixing of the individual cascades
initiated by the initial $q_1\bar{q}_2$ and $q_3\bar{q}_4$ pairs,
which effaces the original identity of the jets, at least in the
central low-momentum (-rapidity) region.
The effect is especially important here
because, as is clear from the discussion in Sec. 3.1,
the bulk of parton emissions occurs at very early times,
while the $W^\pm$ showers are still overlapping, and at
small energy fractions $x$.
Therefore one would expect within our parton-cluster
formation scheme, which is based on the nearest-neighbor separation of a
given pair of partons in conjunction with their color combination, that
locally the number of possible 2-parton combinations would be
significantly increased
to include many possible `exogamous' partnerships between
a parton originating from one 2-jet system and a parton from the
other 2-jet system.
The enhanced  cluster formation possibilities should then
be mirrored in the hadronic mass spectrum,
because the cluster-hadron decays are described as local,
independent processes.
In our analysis, there is no
mechanism for suppressing `exogamy' relative to the normal
assumption of `endogamy', according to
which only partons from the same 2-jet system can
coalesce, with the exception of some limited amount of `color
reconnection' \cite{SK}. We will return to these issues below,
when we present numerical results which confirm these effects.
\medskip

{\bf 3.2.2 Numerical Results}
\smallskip

We have used in our numerical simulations of
$W^+W^-$ production at the nominal LEP2 energy
of 170 GeV the values of the parameters $L_c,L_0, \mu_0$
that we specified in Sec. 3.1, on our discussion of $Z^0$ decays
at LEP1.
To study within our framework whether the spatial overlap of
parton production from the $W^+$ and $W^-$ decays results in
observable effects, we contrast
two extreme sets of initial conditions for the
production of $W^+W^-$ pairs:
\begin{itemize}
\item[a)]
the {\it realistic} case, in which the $W^+$ and $W^-$ decay close by at
space-time vertices $r^\pm$, eq. (\ref{rpm}), with an average separation
$\Delta r < 0.1 \;fm$, and with the spatial
correlations between the two jet systems taken fully into account, and
\item[b)]
a {\it hypothetical} one with the $W^+$ and $W^-$ decay vertices
shifted artificially apart, i.e. $\Delta r \rightarrow  \infty$,
so that spatial correlations between the two jet systems are absent.
\end{itemize}

Fig. 13 exhibits the differences between the
{\it realistic} (full curves) and
the {\it hypothetical} (dotted curves) cases for the cluster mass
spectrum, the transverse momenta of the produced hadrons, and the
multiplicity distribution for charged particles.
In both cases, we show results for the
`color-full' coalescence scenario (III) are shown,
where the effects are most significant.
The main effects of particle correlations in the {\it realistic}
scenario due to the spatial
overlap of the jets during the time-development of the system is
are seen to be (i) a
suppression of cluster production for masses
$m_c\,\lower3pt\hbox{$\buildrel > \over\sim$}\,2$ GeV,
(ii) a softer $p_\perp$-spectrum of resulting hadrons, and
(iii) a significantly shifted multiplicity distribution,
with an enhanced number of
high-multiplicity events and a mean
$\langle n_{ch} \rangle = 37.5$ (30.8 in $|y|<1$),
compared to $\langle n_{ch} \rangle =32.8$ (25.6 in $|y|<1$)
in the {\it hypothetical} scenario.
This shows that multi-particle effects may well not be negligible.
In particular, the difference of more than 10 \% in the
particle multiplicity between the two sets of initial conditions,
which rises to above 20 \% in the central rapidity region $|y|<1$,
should be clearly visible at LEP2. If seen, such an effect would
enable experimentalists to probe this possible
multi-particle situation in more detail, This would in turn
enable our simulation to be improved and extended to more
complicated cases of overlapping cascades, such as appear, for
example, in collisions involving nuclei.
\smallskip

Returning to the three color scenarios (I)-(III) defined in Sec. 2.2,
the next question is the extent to which they differ in their
predictions for the gross features shown in Fig. 13.
Table 3 summarizes for each of the three scenarios
our results for particle multplicities and average event properties,
including the transverse momenta of the
charged particles $\langle p_\perp^{ch}\rangle$, the
charged energy fraction $\langle x_E^{ch}\rangle$ and the
thrust $\langle T\rangle$.
Here $T=\sum_i |p_{\parallel\;i}|/\sum_i |\vec{p}_i|$,
with $\vec{p}_i$ the three-momentum of particle $i$ and
$p_{\parallel\;i}$ its longitudinal momentum
along the axis which maximizes this ratio:
$T$ measures the `jettiness' in the sense that
$T=1$ for a perfect back-to-back configuration and $T=1/2$ for an isotropic
event.
Compared with the analogous Table 1 of Sec. 3.1,
the outcome is, not surprisingly, very similar.
The previously fixed values of $\mu_0$ are the only constraint
on the particle production in each of the color scenarios.
For the reasons discussed before, again the
color-blind scenario (I)
gives the largest parton multiplicity but the smallest
total number of final hadrons, whereas the color-full scenario (III)
features significantly less perturbative parton production, but
the largest total hadron multiplicity.
Table 4 lists some properties of the cluster formation scheme
related to the parton and hadron content.
Compared with the corresponding Table 2 of Sec. 3.1,
one sees that the ratio between partons and clusters, and
between clusters and hadrons, as well as the mean
cluster size and radius,
are roughly the same for both event types (\ref{ee1}) and (\ref{ee2}).
This implies that the formation and hadronization of the clusters
are rather universal and largely
independent of the total energy of the system, a result
that is plausible, given that our approach works locally
in space-time and therefore is insensitive to global
properties of the system.
Concerning the relative contributions of the
various 2-parton cluster formations, as before in Table 1,
most processes in the color-full scenario (III) are of
non-color-singlet types, and hence accompanied by colored
parton emission. For this reason, and since we consider
scenario (III) to be the most complete one, we infer that the
color-singlet scenario (II) may not always be reliable.
Overall however, all three scenarios give rather similar predictions
for these global event properties, so we now proceed to examine
more sensitive quantities.
\medskip

At LEP2, the $W$ mass is one
quantity which is of great physics interest and
which is also potentially among the most sensitive to
particle- and color-correlations.
As already explained, we expect the apparent $W$ mass to
depend in a non-trivial way on the
space-time history of the system,
the color stucture of the evolution, and
on all particle momenta.
It will be an experimental challenge to
measure the $W$ mass accurately using jet
reconstruction. If the measured jets could be correctly separated,
so that each final-state particle could be assigned
unambiguously to the decay of
either the $W^+$ or the $W^-$, then the four-momenta of
$W^+$ and $W^-$ could be reconstructed to give the $W^\pm$ masses.
However, as has been pointed out in Ref. \cite{SK},
in practice such an analysis faces several complications
in addition to purely statistical errors, such as the
removal of background events, mistakes or ambiguities in
assigning individual particles to jets, missing particles and
other problems of a technical nature.
We will not address these here, but discuss in the following
only the physical effects on the $W$ mass determination
due to particle- and color-correlations during the QCD
evolution of the combined $W^+W^-$ system.
In view of the experimental goal at LEP2 of measuring $m_W$ with
$\approx 50$ MeV accuracy,
any physical effect that induces a mass-shift significantly larger than
about 100 MeV, compared to the nominal experimental value of 80.22 GeV, is of
great importance, as would be any apparent
broadening of the Breit-Wigner mass distribution (\ref{breitwigner}).
Such a shift and broadening could be caused by ambiguous
assignments, or even misassignment
\footnote{
The direction of any such
effective mass shift is not obvious {\it a priori}.
However, a `randomization' effect due to experimental
misassignment of particles, i.e. attributing hadrons
to one $W$ that in fact `belong' to the other one, would
presumably cause an upward shift in general.
},
of some of the measured particles between the original $W^+$ and $W^-$.
In particular, within any approach to cluster formation which
allows for `exogamous' coalescence, such as ours, it is not
even possible in principle to assign all final-state hadrons
unambiguously to one $W$ or the other.
Since, as discussed earlier, the spatial overlap between the decay
products of the $W^+$ and $W^-$ is almost perfect, the only
natural scale for the net effect on the
reconstructed average mass $\overline{m}_W = (m^++m^-)/2$
is the QCD energy scale $\Lambda_{QCD}$, which is several hundred
MeV.  To gauge the possible magnitude of this effect, we compare
the {\it realistic} case of overlapping $W$ production and
decay with the {\it hypothetical} case of far-separated $W$
pairs that we introduced earlier in this Subsection.
\medskip

In order to obtain an estimate of the possible change in the
$W$ mass spectrum due to the increased particle density and
the color structure of the parton-cluster conversion, we have
simulated, for each of the three scenarios (I)-(III) of Sec. 2.2, the
experimental jet reconstruction procedure
and the subsequent determination of $m_W$.
We assume that all particles are perfectly measured, so that
our analysis exhibits only the physical smearing of the
separate identities of the $W^+$ and $W^-$ associated
with the correlation effects of
overlapping cascades.
To model the experimental reconstruction of the four
jets initiated by the $W^+$ and $W^-$ decays from the observed
final-state hadrons, we have adopted the
$e^+e^-$ jet-finding algorithm of the JETSET program \cite{pythia},
which is frequently used in experimental analyses.
The general purpose of this algorithm is to determine the
individual jet axes in events with multiple jets ($n_{jet} >2$)
by grouping together measured particles which are near by in phase space,
and then identifying each well-separated group of particles as a
jet with the jet axis given by the constituent particle directions.
Other jet-finding programs exist, such as the JADE \cite{jade}
or the Durham \cite{durham} algorithms, which are conceptually very similar.
Specifically, in the JETSET algorithm a jet is defined as a
collection of particles that have a limited transverse momentum
with respect to a common jet axis, and hence also with respect to
each other. A momentum distance measure $d_{ij}$ is introduced
which essentially measures the relative transverse momentum
of two particles with momenta $\vec{p}_i$ and $\vec{p}_j$,
\begin{equation}
d_{ij} \;:=\;
\frac{4 |\vec{p}_i|^2 |\vec{p}_j|^2 \;\sin^2(\theta_{ij}/2)}
{(|\vec{p}_i|+|\vec{p}_j|)^2 }
\;.
\label{dij}
\end{equation}
A jet is reconstructed by first searching for the highest-momentum
particle $i$, and then all particles $j$ that are within a distance
$d_{ij} < d_{max}$ around it. Here the parameter $d_{max}$
is the transverse momentum `jet resolution power', which depends
on the experimental situation and should be chosen such that
the identities of the jets found are well separated.
In our case, we required a minimum of four jets
to be reconstructed, and have found that a value $d_{max} = 10$ GeV
satisfies this criterion appropriately.
Events that yielded five or more reconstructed jets (about 20 \%)
were ignored in the following analysis.
Furthermore, for the $W$ mass determination we followed the suggestion
of Ref. \cite{SK} and required that each jet has an energy of
at least 20 GeV, and that the angle between any pair of
jets should be larger than 1/2 radian. In agreement with
previous studies, we found that
about 60 \% of the total number of simulated events satisfy
these cuts, so that the reconstruction gives four well-separated jets.

The four jets, corresponding to the initially-produced
$q_1,\bar{q}_2,q_3,\bar{q}_4$ (c.f. Fig.12), may be paired into a
candidate $W$ in three different ways to be associated with
the $W^+$ and $W^-$, so that
each event gives three different values for
the reconstructed mass $\overline{m}_W$.
To select the most likely pair configuration, we
exploit our knowledge of the initial
$q_1,\bar{q}_2,q_3,\bar{q}_4$ configuration and
map the four reconstructed jets one-to-one to the original
partons. Out of the three possible permutations, we then pick the one that
minimizes the product of the invariant
masses between each jet and the parton. Formally the
selection criterion is given by:
\begin{equation}
\Sigma^{(4-jet)} \;:=\;
\min \left[ \frac{}{}
\mbox{Perm}_{i,j,k,l}
\left\{
\frac{}{}
M(i,1)\;M(j,2)\;M(k,3)\;M(l,4)
\right\}
\right]
\label{Cjet1}
\;,
\end{equation}
where
\begin{equation}
M(i,j)\;=\; \sqrt{\left(k_i^{(jet)} \,+ \,k_j^{(q)}\right)^2}
\label{Cjet1a}
\;.
\end{equation}
{}From the total energy-momentum of the jet pairs
selected in this way, one then
obtains the two invariant masses $m^+$ and $m^-$ of the
$W^+$ and $W^-$ and finally the average $\overline{m}_W=(m^++m^-)/2$.

We recall that this standard procedure is formulated entirely in
momentum space, because the experimental measurements of
particle momenta do not yield any explicit
information about the spatial distribution or the time of production.
However, in view of the close relation between the space-time evolution
of the particle system and the emergence of hadrons in momentum
space (c.f. Figs. 9-11 and the discussion in Sec.3.1),
one can presume that the above jet reconstruction scheme gives
a reasonable image of the true jet structure and its underlying
space-time history.
In particular, since the momentum distance $d_{ij}$ of two particles
projects out their relative transverse momentum, and is insensitive
to their longitudinal momenta, this distance measure
provides a local criterion which is independent of rapidity.
\smallskip

Fig. 14 exhibits rather vividly the differences between our
color scenarios (I)-(III) in the jet reconstruction procedure.
We plot the mass spectrum of the four reconstructed jets in each
$W^+W^-$ event, and the quantity
$X:= \sum_i^4 |\vec{p}_i^{\;jet}|/\sum_j^n |\vec{p}_j^{\;par}|$,
measuring the fraction of summed jet three-momenta relative to the
total three-momentum carried by the whole ensemble of final particles.
For both quantities, the color-blind scenario (I), which
is solely based on the space-time structure,  deviates
clearly from the scenarios (II) and (III), which in addition take into
account the color structure in the parton-cluster formation
processes. As compared to the latter two,
which give relatively similar results,
scenario (I) is characterized by jets that have, typically, lower mass
and  larger $\langle X \rangle$.

In Fig. 15 we show our final result for the
simulation of the experimental $W$ mass reconstruction. The top part
shows the distribution of the reconstructed $m_W$ for each
of the color scenarios (I)-(III), in comparison to the
$W$ mass distribution which was initially generated
and corresponds to the Breit-Wigner spectrum (\ref{breitwigner}).
We remark that the average value of the latter is
$\langle m_W^{(generated)} \rangle = 79.75$ GeV, and hence below
the nominal `on-shell' value 80.22 GeV, because of the
kinematic competition between the Breit-Wigner line shapes
of the $W^+$ and $W^-$ and the phase-space available for production of the
pair.
As in Fig. 14, the color-blind scenario (I)
sticks out significantly also in Fig. 15,
whereas the color-singlet and color full scenarios,
(II) and (III) give almost identical results.
The lower part of Fig. 15 amplifies this effect,
by showing the effective mass shift
\begin{equation}
\Delta m_W \;=\;
m_W^{(generated)} \;-\; m_W^{(reconstructed)}
\;.
\label{ms0}
\end{equation}
The mean values of the distributions are
\begin{equation}
\langle \,\Delta m_W \,\rangle^{(i)}_{real}\;=\;
\left\{
\begin{array}{cl}
+ 383 \;\mbox{MeV} & \mbox{for $i=$ I} \\
+ 812 \;\mbox{MeV} & \mbox{for $i=$II} \\
+ 706 \;\mbox{MeV} & \mbox{for $i=$III}
\end{array}
\right.
\label{ms1}
\;.
\end{equation}
where the subscript `real' stands for the
{\it realistic} case of spatially-overlapping $W^\pm$ decays, as defined above.
The  widths associated with the $\Delta m_W$-distributions are
large: $\sigma(\Delta m_W) = 2.65/3.05/3.02$ GeV for
scenarios (I)/(II)/(III), but the effect of the wide
tails is presumably not of significance, since any fine-tuned
experimental $W$ mass analysis would be mostly sensitive to the peak.

A large shift $\langle\Delta m_W\rangle_{real}$ would not
be a problem for measuring the $W$ mass if it were independent of
the hadronization model. However, a notable aspect of (\ref{ms1}) is
the difference in mass shift among the three
color scenarios. Defining $\delta m_W^{(II, III)}$
as the difference between the scenarios which
incorporate some color-accounting, namely (II) and (III),
and the scenario without color, namely (I), this relative mass shift
reflects the net effect of particle correlations due to color structure.
It turns out to be impressively large:
\begin{equation}
\langle \,\delta m_W\,\rangle^{(i)}_{real}\;\equiv\;
\langle \,\Delta m_W\,\rangle^{(i)}_{real}\;-\;
\langle \,\Delta m_W\,\rangle^{(I)}_{real}
\;=\;
\left\{
\begin{array}{cl}
 429 \;\mbox{MeV} & \mbox{for $i=$II} \\
 323 \;\mbox{MeV} & \mbox{for $i=$III} \\
\end{array}
\right.
\;.
\label{ms2}
\end{equation}
The differences between the three color scenarios are certainly
not negligible compared with the experimental
goal of measuring $m_W$  with an accuracy of about 50 MeV.
Even the difference between our favored `color-full' scenario (III)
and the next-best scenario (II) in (\ref{ms2}),
\begin{equation}
\langle\delta m_W \rangle^{(III)}_{real}-\langle \delta m_W
\rangle^{(II)}_{real}
\;\simeq\; -100 \;\,\mbox{MeV}
\;,
\label{ms2a}
\end{equation}
is large compared with this experimental goal.

The mass shifts in (\ref{ms1}) and (\ref{ms2}) include a mixture of two
effects:
first,the misassignment of particles to one $W$ that belong in reality to
the other, and second, the forced assignment of hadrons to one $W$ that
emerged from clusters formed by `exogamous' pairs of partons
with mixed origins from the two overlapping $W$ decays.
Some information on the relative contribution of these independent effects
can be opbtained by comparing the $W$ mass shift in the
above {\it realistic} case with very near-by
$W^\pm$ decay vertices (where both misassignments and `exogamous'
clustering are possible), and the
{\it hypothetical} situation with infinitely separated $W^\pm$ decay vertices
(where `exogamy' is impossible, but misassignments are still possible).
Experimentally, the hypothetical case of widely-separated $W^\pm$ decays can
be simulated by superposing the decays of a pair of Lorentz-boosted
$Z^0$ decays measured at LEP1, which should enable
the {\it hypothetical} mass shift
$\langle\Delta m_W \rangle_{hypo}$ to be understood with adequate precision
($\sim 10$ MeV).
It should be noted, though, that misassignment may occur at different rates
in the {\it realistic} and {\it hypothetical} cases, because
of the differences in the final hadron spectra even for `endogamous'
production.
We find in the {\it hypothetical} case the mass shifts
\begin{equation}
\langle \,\Delta m_W \,\rangle^{(i)}_{hypo}\;=\;
\left\{
\begin{array}{cl}
+ 396 \;\mbox{MeV} & \mbox{for $i=$ I} \\
+ 781 \;\mbox{MeV} & \mbox{for $i=$II} \\
+ 424 \;\mbox{MeV} & \mbox{for $i=$III}
\end{array}
\right.
\label{ms3}
\;,
\end{equation}
compared to the mass shifts (\ref{ms1}) in the {\it realistic} situation.
Table 5 gives a comparative list of the various mass shifts in the different
cases.
Taking the differences, we infer that hadrons produced `exogamously'
are directly responsible for shifts,
\begin{equation}
\langle \,\Delta m_W \,\rangle^{(i)}_{real}\;-\;
\langle \,\Delta m_W \,\rangle^{(i)}_{hypo}\;=\;
\left\{
\begin{array}{cl}
-  13 \;\mbox{MeV} & \mbox{for $i=$ I} \\
+   6 \;\mbox{MeV} & \mbox{for $i=$II} \\
+ 282 \;\mbox{MeV} & \mbox{for $i=$III}
\end{array}
\right.
\label{ms4}
\;.
\end{equation}
In the absence of a more appropriate comparison, we can only regard
the difference between the last two entries in (\ref{ms4}),
provisionally, as a conservative estimate
\footnote{
We find qualitatively similar numbers for the mass shifts if we select
events with dijet masses in the nighborhood of $m_W$.
}
of the error in determining $m_W$ from purely-hadronic $W^+W^-$ final states.

It should be emphasized that, in the discussion
of the previous paragraph, we have
been incorporating theoretical knowledge not available to experiments at LEP2,
namely the energy-momentum vectors of the initially-produced partons
$q_1,\bar{q}_2,q_3,\bar{q}_4$ configuration, which we exploited to
remove the jet assignment ambiguity via the criterion (\ref{Cjet1}).
In reality this knowledge is absent, so that some other criterion must be found
for assigning the four observed jets to the $W^\pm$.
A systematic study of possible criteria
for assigning the four observed jets to the $W^\pm$ and $W^-$
involves experimental detector characteristics, and lies beyond the
scope of this paper.
However, to assure the reader that our results are not dependent on the details
of the reconstruction method, but in fact reflect the underlying
simulated particle dynamics, we repeated the $W$ mass analysis
using an alternative criterion that makes no {\it a priori}
use of knowledge of the initial parton configuration as (\ref{Cjet1}), but
solely relies on the final particle momenta. Instead of (\ref{Cjet1})
we picked out of the three possible pair configuration of the
four reconstructed jets the one that minimizes the
difference between the jet masses and the nominal $W$ mass value
$m_W = 80.22$ GeV that we introduced before:
\begin{equation}
\widetilde{\Sigma}^{(4-jet)} \;:=\;
\min \left[ \frac{}{}
\mbox{Perm}_{i,j,k,l}
\left\{
\frac{}{}
\left| \widetilde{M}(i,j)\,-\, m_W \right| \;+\;
\left| \widetilde{M}(k,l)\,-\, m_W \right|
\right\}
\right]
\label{Cjet2}
\;,
\end{equation}
where
\begin{equation}
\widetilde{M}(i,j)\;=\; \sqrt{\left(k_i^{(jet)} \,+ \,k_j^{(jet)}\right)^2}
\label{Cjet2a}
\;.
\end{equation}
Using this more experimental reconstruction scheme, we obtain results very
similar
to those in (\ref{ms1}) and (\ref{ms2}).
Table 5 compares the outcome of the two different
reconstruction schemes (\ref{Cjet1}) and (\ref{Cjet2}).
{}From this it is evident, that the latter, more experimental scheme
gives results with a spread similar to those in (\ref{ms4}) above.
Further study of the magnitude of the particle- and color-correlations
will require more detailed experimental simulations.
\smallskip

Although we believe that the `color-full' scenario (III) is the most reliable,
prudence demands that one consider the differences between
this and the other scenarios as estimates of the possible
systematic error in the determination of $m_W$, until deeper
understanding of the model-dependence of the mass determination
is available.
For example, one issue, which lies beyond the scope of this paper,
is the effect of Bose-Einstein correlations \cite{SK,LS}.
An alternative way of thinking about the above
results would be to take the value of $m_W$ from elsewhere
(e.g., from $W^+W^-\rightarrow l \,\nu\, q\bar q$), and
regard hadronic $W$ pairs as a laboratory for studying
the effects of space-time and color correlations in a
controlled environment. However, we are reluctant to retreat to
this position until more model studies and comparisons have been
made. We comment on these questions in the final section of
this paper, discussing in particular why
our result for the shift $\langle \delta m_W\rangle$ is
up to an order of magnitude larger than the previous
estimates of mass shifts due to color correlations within
the `color reconnection' approach \cite{SK,GH,BW,LL,sharka}.
\bigskip
\bigskip

\noindent {\bf 4. DISCUSSION AND COMPARISON WITH PREVIOUS WORK}
\bigskip

In this work we have focussed on two issues of fundamental interest
for general high-energy particle collisions in QCD:
first, the space-time development of
perturbative parton cascade evolution
and its interplay with the non-perturbative hadronization dynamics,
and second, the issue of color correlations among
the partons during the evolution, their impact on hadron formation,
and their possible observable consequences for experiments.
In view of the upcoming LEP2 experiments at CERN, we studied
specifically $e^+e^-$-collisions at
$\sqrt{s}= 91$ GeV and $\sqrt{s}= 170$ GeV, corresponding to the energies
at LEP1 and LEP2, respectively.
We emphasize however, that the issues addressed here are of
general relevance to,
e.g., HERA physics (both in $ep$ and $eA$ collisions) and
Tevatron experiments ($\bar p p$ collisions), and experiments
at RHIC ($AA$ collisions) and the LHC ( $pp$, $pA$ and $AA$
collisions).

Our analysis has been in the context of
the specific approach that we have developed for
describing the time evolution of a generic mixed parton/hadron system
in both position and momentum space.
The essential novel features of our approach are to combine
kinetic theory techniques with
the well-established perturbative parton evolution in momentum
space, and the use of
a phenomenological model for parton-cluster recombination
which is based on spatial separation as a criterion for
confinement.
Our kinetic description
allows us to follow the
time evolution of the system locally
in each phase-space element $d^3rd^4k$ in a manner
consistent with the uncertainty principle, and
trace the evolution from the initial state all the way
to the final hadron yield. Thus we ``see'' the time-dependence
from the ``hot'' initial state as far as possible as the
participating partons and hadrons experience it
at the microscopic level.
There is certainly considerable
model dependence associated with our phenomenological
simulation of the hadronization mechanism, but
the comparisons here and in Ref. \cite{ms37} with
experimental data of global event properties measured at LEP1
indicate that our approach is largely consistent
with the present knowledge of LEP physics.

The main findings from our numerical simulations are:
\begin{itemize}
\item[(i)]
The space-time development of jet evolution exhibits the
characteristic features of an `inside-outside' cascade
\cite{bj,kogut73}, in which particles are in the average produced earliest
at smallest rapidity (small $x$), close to the spatial location
of the initial vertex.
Only with progressing time and increasing distances do
particles populate more densely regions of higher rapidity, as
we discuss in more detail below.
\item[(ii)]
Color correlations are important, as seen from our comparison of
three scenarios with different levels of accounting for the
color degrees of freedom.
These correlations occur among the partons during both the perturbative
parton shower and the coalescence of partons into clusters, exhibiting
themselves, e.g., in an increase of high-multiplicity events, larger
average hadron multiplicities, and softer momentum spectra for the
produced particles.
\item[(iii)]
Our simulation of experimental jet reconstruction from the
final-state hadrons and the extraction of the derived
$W$ mass spectrum raises the possibility that there may be
a large upward shift in the `observed'
$m_W^{(reconstructed)}$ from the `actual' mass $m_W^{(generated)}$,
which could reach several hundred MeV.
The effect of color correlations alone (\ref{ms4}) could be as large as
$\langle \Delta m_W\rangle\approx 300$ MeV, which is up to
an order of magnitude larger than estimated in
similar previous investigations \cite{SK,GH,BW,LL}.
If true, this would have dramatic implications for the
LEP2 experiments.
\end{itemize}

We must then ask why the space-time and
color-correlation effects turn out to be so large in our
analysis, compared to the related investigations by, e.g.,
Sj\"ostrand and Khoze (SK) \cite{SK}, and by Webber (BW) \cite{BW},
who find in general that
$\langle \delta m_W\rangle \approx 20-40$ MeV, rising possibly
to $\approx 100$ MeV in extreme cases.
The answer is likely to be rooted in at least two
essential differences between the approaches
employed by SK and BW, and ours. As we have already tried to
bring out, our approach incorporates an explicit causally-ordered
space-time description, and a dynamical scheme of parton-hadron
conversion that is formulated spatially.
At this point, it is important to understand better these differences
as an aid to judging the relevance of their possible
implications.

The common basic concept of SK and BW is the use of well-understood
angular-ordered parton shower evolution according to the
leading-logarithmic QCD evolution equations in momentum space,
i.e., in terms of $x$, and $k^2$ or $k_\perp^2$,
in combination with a specific hadronization model,
string-fragmentation in the case of SK and cluster-fragmentation in
the case of BW. The perturbative parton evolution and the
non-perturbative hadronization are treated independently,
with an interface at some intermediate mass scale $\mu$ which
plays the role of a perturbative cut-off parameter.
No explicit use of space-time variables is made {\it a priori},
but the space-time location of each parton at the moment of
hadronization can be reconstructed statistically by working
backwards. This procedure does not, however, follow causally
the space-time histories of all the partons including their
space-time correlations, which is accomplished naturally by our
forward time evolution approach.
In the previous approaches, interference and cross-talk between
the parton showers emanating from the $W^+$ and $W^-$ decays
are incorporated by allowing for the
color reconnection of partons belonging to
different color-singlet subsystems
just before they are hadronized. At most one such color reconnection
is allowed to occur in each event, i.e., out of the ensemble
of final state partons only a single pair is possibly selected to be
rearranged in color space. The reconnection probability
which controls the strength of the effect is treated as a free
parameter.

The central intuition of our approach, as applied to $e^+e^-$
collisions, is that the intial state may be regarded as a `hot spot'
where partons are temporarily unconfined and evolve according to
perturbative QCD, surrounded by a `cold' confining medium
containing non-perturbative condensates, through which hadrons
propagate. During perturbative QCD shower evolution, the hot
spot expands stochastically and inhomogeneously. We follow the
space-time evolution in small discrete time steps, monitoring
the separations between all the partons. In this way, we keep
track of all their spatial correlations during the entire
shower evolution. In the particular case of $e^+e^- \rightarrow
W^+W^-$, it is easy to see, as we discussed earlier, that the
$W^+$ and $W^-$ decay while still very close to each other and
generally within a typical confinement length of about $10^{-13}$ cm,
so that they together occupy a single `hot spot'. We ignore the
$W^+$ and $W^-$ `parentages' of the partons during the subsequent
combined shower evolution. The likelihood of
non-perturbative hadronization (confinement) is determined
statistically by the nearest-neighbor criterion (\ref{L}) alone,
without regard to the parentage of the individual partons.
If the $W^+$ and $W^-$ showers were to overlap completely, the
probabilities of `exogamous' cluster formation by partons with
different $W^{\pm}$ parents would be comparable with that of `endogamous'
cluster formation by partons sharing the same $W^{\pm}$ parent.
This would lead to many more `exogamous' clusters than the
color-reconnection models proposed previously. In practice,
the showers from the $W^+$ and $W^-$ decays usually have different
axes, and phase mismatches between high-rapidity partons in any
case suppress their possible interferences, so that `exogamy' is
likely only between partons with low momenta in the center-of-mass.
Nevertheless, there are many such partons, and their overlap is
large, so there are many opportunities for `exogamy', as we now
discuss more quantitatively.
\smallskip

We recall that when $W^+W^-$ pairs are produced at
$Q=\sqrt{s}=170$ GeV in the center-of-mass frame of the $\gamma^\ast/Z^0$,
the $W^{\pm}$ are to
first approximation created at rest, because their boost factors are
$\gamma^\pm \simeq Q/(2m_W) = 1.05 \approx 1$.
As pointed out in 3.1.1, the difference $\Delta t$
between the times at which the
$W^+$ and $W^-$ decay is short on a typical hadronic scale:
$\Delta t = |t^+-t^-| \simeq 0.1$ $fm$. Suppose, for definiteness,
that the $W^+$ decays first.
Then, for a time span $\Delta t$ the
produced $q_1\bar{q}_2$ system evolves undisturbed,
just as in ordinary 2-jet events.
A parton shower develops, spreading rapidly out in configuration
space, analogously to the illustration in Fig. 6.
In space-time, the $q_1\bar{q}_2$ pair
and the increasing number of accompanying bremsstrahlung partons expand
like in a shock wave \cite{ms37} from the decay vertex. Most of
the activity is concentrated in cones around the leading
$q_1$ and $\bar{q}_2$. The situation is illustrated in Fig. 16, where
we define the initial 2-jet orientation as the $z^+$-axis.
Although the
particles commonly move outward close to the speed of light
with velocities determined by their invariant mass $k^2$, the
the region of space-time extent that they occupy
depends sensitively on the energy fractions $x$
via the uncertainty principle.
As a consequence, the
leading partons ($x =O(1)$) become separated from the
wee partons ($x \,\lower3pt\hbox{$\buildrel < \over\sim$}\,10^{-2}$).
This is an important effect of time dilation and
Lorentz contraction \cite{bj,kogut73}.
In fact, the $q_1$ and $\bar{q}_2$ moving along the $z^+$-axis
form the forefront of the expanding shell, which has a proper
thickness $\Delta z^{proper} \simeq 1/m_W$, implying
\begin{equation}
\Delta z_{q} \;=\; \Delta z_{\bar q} \;\equiv \;
\frac{\Delta z^{proper}}{2\,\gamma^+}\;
\simeq\;\frac{1}{2\, \gamma^+\,E^+}\;\approx\; 10^{-3}\;fm
\;,
\label{zqq}
\end{equation}
whereas the wee partons, in particular the softest ones which are
emitted earliest by the $q_1$ and $\bar{q}_2$, are smeared out
by the uncertainty principle
over a comparably large longitudinal extent,
\begin{equation}
\Delta z_{wee} \;\equiv\;\frac{1}{2\,(k_z)_{wee}}\;
\simeq\;\frac{1}{2\,x\;E^+}\;\approx\; 0.1 - 1\;fm
\;\;\;\;\;\;\;\;(\mbox{for $x \approx 10^{-2}-10^{-3}$})
\;.
\label{zwee}
\end{equation}
This spatial uncertainty associated with
the particles' momenta implies
that the low-momentum quanta are not all
concentrated at the shock front,
as one would naively expect for massless particles, but
rather occupy the whole interior of the expanding volume,
with the wee partons piling up around the center, as
depicted graphically in Fig. 16.

In this environment the $W^-$ now decays
about 0.1 $fm$ after the showering of the
$q_1\bar{q}_2$ system from the $W^+$ has already produced a
significant number (of the order of 10) wee partons around
its vertex at $\vec{r}^{\;+}$, producing
a second $q_3\bar{q}_4$ pair
which expands around $\vec{r}^{\;-}$ in the same fashion.
The second jet system is produced within the wee-parton cloud
of the first one, which had been created {\it in vacuo}.
Therefore, as suggested by Fig. 17, these later partons
experience a `medium' of surrounding wee partons, immediately
opening up the possibility of interactions due to correlations.
The leading $q_3\bar{q}_4$ (plus possibly a few large-$x$ partons)
of the second jet system are well separated in rapidity
$y\simeq \ln x$ from the small-$x$ wee partons, and therefore
decouple and escape essentially unscathed \cite{khoze94}.
On the other hand, the wee partons produced in $W^+$ decay
and those emitted by the later $W^-$ decay
can interact easily, because they are in the same range of $y$ ($x$)
and moreover overlap in their spread of spatial directions
(\ref{zwee}).

These expectations are borne out by Fig. 18, which illustrates
the fractions of clusters that are formed `endogamously', i.e., by
partons from the same W decay shower, and `exogamously', i.e., by
interactions between partons from different W showers. The clusters
formed by endogamous unions of partons
from the $W^{\pm}$ are labelled by ${\cal E} =0, 1$, respectively,
and clusters formed by exogamous union between partons from
the $W^+$ and the $W^-$ are labelled by ${\cal E} = 1/2$. In some cases, the
partons emitted during cluster formation combine to form other
clusters with ${\cal E} = 1/4, 3/4$, etc.. We see in Fig.~18 that there
are in general several exogamous unions per event,
and that these are more
common among the less-energetic clusters, as suggested qualitatively
in the previous paragraph.
(For comparison, in the hypothetical case of inifinitely separated
$W^+$ and $W^-$ decay vertices, where `exogamy' is excluded, the distributions
would be peaked at 0 and 1, and vanishing in between.)
We notice also that our our preferred
{\it Color-Full} scenario III produces the largest fraction
of exogamous clusters, as
one would expect qualitatively, in view of the larger number of
allowed diagrams shown in Fig.~5. In our interpretation, it is the
large number of these exogamous unions that is primarily responsible
for the large mass shift in eq.~(40).
\smallskip

We believe that these intuitive arguments provide a
plausible origin for the
large non-perturbative correlation effects found in our approach.
We would like to emphasize that the space-time picture outlined
above is, in its general features,
independent of the particular Lorentz frame employed,
as has been discussed especially by Bjorken \cite{bj} and by
Kogut and Susskind \cite{kogut73}.
A Lorentz boost between different frames can distort the picture,
but the characteristic `inside-outside' evolution in both rapidity and
longitudinal direction is always the same.
We think that any such `inside-outside' space-time picture
will lead to considerable overlap, correlation and interaction
among the wee partons from the $W^+$ and the $W^-$ (c.f. Fig 17).
These effects are subject to the general constraints of
relativity and the uncertainty principle, but
one must make explicit use
of space-time variables if one wishes to model them realistically.
In our approach the causal evolution
is followed in a probabilistic manner, with
the phase-space density of particles at any point of time being
determined by the preceding history, and in turn governing the
statistical occurrence of parton-cluster
coalescence, according to the local particle density and the
nearest-neighbor criterion (\ref{L}).
On the other hand, a pure momentum-space description
cannot take correlation effects fully into account,
which may lead to
a substantial underestimate of the possible effects on
observable quantities such as the $W$ mass.

Some final comments are appropriate here.
It has to be stressed that the systematic shift $\delta m_W$
that we find in the $W$ mass is dependent on
our specific modelling of the non-perturbative
parton-cluster formation, as can be seen from the fact that it
depends on the color scenarios (I)-(III) which are used. We recall
that we have not considered color rearrangement on the
perturbative level, and that its
effects have been estimated to be very small,
$\delta m_W^{pert.}\,\lower3pt\hbox{$\buildrel < \over\sim$}\,5$
MeV \cite{SK}, and not comparable with
our results (\ref{ms2}),
$\delta m_W \simeq 300 - 400$ MeV.
As we have emphasized repeatedly,
the non-perturbative aspects have been modelled
rather differently in SK and BW, and we cannot prove that
our results are more reliable.
However, we believe that our results demonstrate
that the effects of color flow dynamics during
the non-perturbative parton-hadron conversion are model-dependent,
and that one must be cautious in drawing conclusions from
an incomplete set of different approaches.
It will be possible to test models for $W^+W^-$ hadronization using
measurements of final-state hadron distributions in both
longitudinal and transverse momenta for different relative orientations of
the $W^\pm$ decay jets.
This may enable the development of strategies for the measurement
of $m_W$ that are insensitive to complexities associated with
low-momentum hadrons that have been the central theme of this paper.
It remains to be seen what the experiments at LEP2 will soon reveal.
\smallskip

We close by emphasizing  that the effects discussed here become of increasing
importance as one considers more complex systems
with larger particle densities: perhaps already in deep-inelastic
lepton-hadron scattering or
hadronic collisions at very small $x$, but certainly in reactions
involving nuclei, such as $eA$ or $pA$, where partons
propagate through and interact with nuclear matter, and especially in $AA$
where multiple (mini)jets (up to many hundreds for large $A$)
evolve simultanously and interact.
\bigskip
\bigskip

\noindent {\bf ACKNOWLEDGEMENTS}
\bigskip

We thank Valery Khoze and Yuri Dokshitzer for encouraging discussions,
and appreciate useful suggestions by Leif L\"onnblad.
Also we thank Brookhaven National Laboratory for generously providing
computer time on its RHIC-cluster.

\bigskip
\bigskip
\bigskip

\newpage

{\bf TABLE CAPTIONS}
\bigskip

\noindent {\bf Table 1:}

Average multiplicities of produced partons and hadrons from
simulations of $e^+e^- \rightarrow Z^0 \rightarrow hadrons$
at $Q=\sqrt{s}=91$ GeV. For each of the color scenarios
(I)-(III), the experimental number for the total
charged multiplicity was used to fix the
minimum parton virtuality $\mu_0$ during the perturbative
shower activity.
With this constraint, all other numbers emerge as predictions.
\bigskip

\noindent {\bf Table 2:}

Results for the color scenarios (I)-(III) of average
multiplicity, size and mass of
clusters formed from coalescing partons in
$e^+e^- \rightarrow Z^0 \rightarrow hadrons$
at 91 GeV, as well as the relative contributions of the
different cluster formation subprocesses of Fig. 5.
\bigskip

\noindent {\bf Table 3:}

Listing of average properties of
$e^+e^- \rightarrow \gamma^\ast / Z^0 \rightarrow W^+W^-\rightarrow hadrons$
at $Q=\sqrt{s}= 170$ GeV resulting from the simulations
for the color scenarios (I)-(III).
The same values for the minimum parton virtuality $\mu_0$
as in Table 1 were used.
Mean multiplicities of partons, hadrons, and charged particles
are shown,
as well as the averages of charged particle transverse momentum
$p_\perp^{ch}$ (with respect to the thrust axis),
energy fraction $x_E^{ch}=2E^{ch}/Q$, and thrust $T$.
\bigskip

\noindent {\bf Table 4:}

Average cluster properties in
$e^+e^- \rightarrow \gamma^\ast / Z^0 \rightarrow W^+W^-\rightarrow hadrons$
at 170 GeV for each of the color scenarios (I)-(III).
In correspondence to Table 2,
the average multiplicity, size and mass of
clusters are listed, and
the associated relative contributions of the
different cluster formation subprocesses of Fig. 5.
\bigskip

\noindent {\bf Table 5:}

Compilation of the mass shifts resulting from the various investigations
discussed in the text.
As before, (I), (II) and (III) refer to the `color-blind',
`color-singlet', and `color-full' scenario, respectively.
The subscripts `real' and `hypo' label the
two discussed extremes of the
{\it realistic} situation (expected close-by decays of the $W^+$ and $W^-$),
and a {\it hypothetical} case (with infinitely-separated
$W^+$ and $W^-$ decay vertices).
The top part refers to the jet-selection scheme (\ref{Cjet1}),
and we compare the reconstructed average mass
$\langle m_W^{(rec.)} \rangle =   \langle m^+ + m^- \rangle / 2$,
the mass shifts
$\langle \Delta m_W \rangle = \langle m_W^{(gen.)} \rangle - \langle
m_W^{(rec.)} \rangle$,
where the mean generated $W$ mass is
$\langle m_W^{(gen.)} \rangle =    79.75$ GeV,
and the widths $\sigma(\delta m_W)$ of the associated distributions (Fig. 15).
The lower part of the table presents the corresponding
quantities extracted by using the jet-selection scheme (\ref{Cjet2}).

\newpage

{\bf FIGURE CAPTIONS}
\bigskip

\noindent {\bf Figure 1:}

Illustration of the dynamical connection between the three
basic process types described by the equations (\ref{e1})-(\ref{e3}):
a cascade develops first as a shower with parton multiplication,
followed by coalescence of partons to color-singlet clusters, and finally
the decay of clusters into hadrons.
\bigskip

\noindent {\bf Figure 2:}
Schematics of the {\it nearest-neighbor criterion}, eq. (\ref{L}).
Two partons may coalesce to a cluster if they are nearest neighbors
in space-time, and if their mutual separation
$L_{12}$ approaches the confinement scale $L_c$, with a probability given
by the width of the transition interval $[L_0,L_c]$ (c.f. eq. (\ref{Pi3})),
which originates from the form of the confinement potential
$V$ in (\ref{Lagrangian}).
\bigskip

\noindent {\bf Figure 3:}

Example of color-singlet combinations of partons in a quark
jet which form independent
subsystems, and of color-connected partons which screen the color of the
original quark.
\bigskip

\noindent {\bf Figure 4:}

Diagrammatic rules for the color flow structure in
parton branching processes, connecting
quarks and gluons coming in and going out of a vertex. Each quark (antiquark)
is accompanied in its direction of motion by a color (anticolor) line,
and each gluon carries both a color and an anticolor line.
\bigskip

\noindent {\bf Figure 5:}

Diagrammatic rules for the color flow structure in
the various cluster formation processes.
Each quark (antiquark) carries a color (anticolor) line,
and each gluon carries both a color and an anticolor line.
Note that the restriction to color-singlet configurations
of two coalescing partons is equivalent to considering
only the three diagrams in the left column. The remaining
four diagrams require additional parton emission to ensure
local color conservation.
\newpage

\noindent {\bf Figure 6:}

Illustration of the space-time development of the parton shower,
cluster formation by parton coalescence,
and cluster decay into hadrons in an
$e^+e^-\rightarrow Z^0 \rightarrow q\bar q$ event.
The time intervals for the propagation of intermediate particles
are determined by the particles' 4-momenta and the uncertainty principle.
\bigskip

\noindent {\bf Figure 7:}

Time development and conversion from parton to hadron degrees of freedom
in $e^+e^- \rightarrow hadrons$ at $Q=\sqrt{s}=91$ GeV in the
center-of-mass frame.
We show the time dependences of the total number of
`live' partons (hadrons),
their respective total energies $E=\sum_i E_i$, total longitudinal
momentum fraction $x = 2/Q \sum_i k_{z\;i}$, and their summed
transverse momentum $k_\perp = \sum_i \sqrt{k_{x\;i}^2 + k_{y\;i}^2}$,
where the $z$-axis is defined along the jet direction.
The three curves correspond to the three color scenarios (I)-(III)
defined in the text.
\bigskip

\noindent {\bf Figure 8:}

Effects of the three color scenarios (I)-(III) on
cluster size distribution and cluster mass spectrum (top),
and on the rapidity and transverse momentum distribution of
produced hadrons (bottom), in $e^+e^-\rightarrow hadrons$
at 91 GeV.
\bigskip

\noindent {\bf Figure 9:}

Time development of momentum and spatial
distribution of partons present in the system at given times.
The top parts show rapidity and transverse-momentum spectra, whereas
the bottom plots show
longitudinal and transverse spatial distributions, where the $z$-axis is
defined along the jet direction.
The various curves correpond to different time steps during the evolution
as specified in the top left corner.
To guide the eye, the arrows indicate the direction of time change.
\bigskip

\noindent {\bf Figure 10:}
Time development of hadron distributions in correspondence
to the parton spectra of Fig. 9.
The different curves show the contributions at the time steps
given in the top left corner.
\bigskip

\noindent {\bf Figure 11:}

Two-dimensional plots of the ratios of hadrons to the sum over
partons plus hadrons,
eq. (\ref{phratio}). The top part shows the spectrum in rapidity
versus $\log (t)$, where $t$ is the time in the center-of-mass frame.
The bottom part shows the corresponding distribution along
the jet ($z$-) axis.  The `inside-outside' evolution in both
$y$ and $z$ with increasing time is clearly seen.
\bigskip

\noindent {\bf Figure 12:}

General situation of initial 4-jet production in
$e^+e^- \rightarrow \gamma^\ast / Z^0 \rightarrow W^+W^- \rightarrow hadrons$.
The $W^+$ and $W^-$ are simultanously produced
at $t_0$ by the $\gamma^\ast/Z^0$, and then decay into
jet pairs $q_1\bar{q}_2$ and $q_3\bar{q}_4$, at delayed times
$t^+$ and $t^-$ with
$|t^+-t^-|\,\lower3pt\hbox{$\buildrel < \over\sim$}\,0.1$ $fm$.
\bigskip

\noindent {\bf Figure 13:}

Results from the {\it realistic} (full curves) and
the {\it hypothetical} (dotted curves) evolution of
the $W^+W^-$ system defined in the text.
We show the cluster mass spectrum, transverse momentum of produced hadrons, and
multiplicity distribution of charged particles.
\bigskip

\noindent {\bf Figure 14:}

Results for the
color scenarios (I)-(III) from the simulated reconstruction
of the $W^+W^-$ 4-jet system according to the
jet-finding algorithm explained in the text.
The top part shows
the mass spectrum of the reconstructed four jets (before
imposing cuts on $m_{jet}$), whereas the bottom part
shows the fraction of the total three-momentum of particles carried
by the four reconstructed jets:
$X:= \sum_i^4 |\vec{p}_i^{\;jet}|/\sum_j^n |\vec{p}_j^{\;par}|$.
\bigskip

\noindent {\bf Figure 15:}

Results of the
simulation of experimental $W$ mass reconstruction for the
three color scenarios (I)-(III). The top part
shows the distribution of the reconstructed $m_W$
and, as reference, the generated $W$ mass distribution
from the Breit-Wigner spectrum.
The bottom part shows the resulting mass shift
distribution
$\Delta m_W = m_W^{(generated)}- m_W^{(reconstructed)}$.
with mean values given by (\ref{ms1}).
\bigskip

\noindent {\bf Figure 16:}

Graphical illustration of the space-time geography
of  jet evolution (part 1):
on the left, the situation of the
$q_1\bar{q}_2$ jet system produced in empty space by  the first decaying
$W$ (here the $W^+$).
In accordance with the `inside-outside' cascade picture, the small-$x$
wee partons are emitted the earliest and occupy the largest longitudinal
spatial region due to the uncertainty associated with their momentum.
They form the bulk of radiated particles.
The leading quark (antiquark)  and the most energetic large-$x$ partons,
on the other hand,  are localized
around the expanding shock front of the system.
The right part shows the same situation in the familiar
longitudinal $t-z$ plane, where the leading particles
separate along the light cone,
and the wee partons fill the central region. Here $t^+$ refers to the
time in the $q_1\bar{q}_2$ rest-frame, and $z^+$ to the jet axis.
\bigskip

\noindent {\bf Figure 17:}

Graphical illustration of the space-time geography
of  jet evolution (part 2):
the left side shows the situation now
for both jet pairs,
$q_1\bar{q}_2$ from the $W^+$ and
$q_3\bar{q}_4$ from the $W^-$.
Due to the large spatial smearing of the small-$x$ wee partons,
there is a significant region of overlap in which interactions
among the partons from the two jet systems can cause correlations.
The right side illustrates as in a) the situation in the
$t-z$ plane, where $t^+\;(t^-)$ and $z^+\;(z^-)$ correspond to the
time and the jet-axis of the
$q_1\bar{q}_2$ ($q_3\bar{q}_4$) jet pair system.
Because at $\sqrt{s} = 170$ GeV the Lorentz factors
$\gamma^\pm \simeq 1$, one has $t^+\simeq t^-$ so that
$t \equiv (t^++t^-)/2$ approximately measures the common time.
Note also that
due to the relative azimuthal angle of the 2 jet systems, the
$z^+$ and $z^-$ axes are generally rotated, so that the
indicated region of wee parton overlap is generally asymmetric.
\bigskip

\noindent {\bf Figure 18:}

Normalized distributions of the
`endogamous$/$exogamous' parentage of clusters formed
in $W^+W^-$ events at $\sqrt{s} = 170$ GeV.
The shower-initiating $q_1\bar{q}_2$ ($q_3\bar{q}_4$)
pairs from the $W^+$ ($W^-$) decay carry
${\cal E} =0$ (${\cal E} =1$). In branching processes,
each endogamous daughter
parton $i$ carries the previous generation's value,
i.e. ${\cal E}_i = {\cal E}_{i-1}$, and
each cluster  formed by exogamous coalescence of partons $i$ and $j$ carries
${\cal E}_c =({\cal E}_i+{\cal E}_j)/2$.
The figure compares the ${\cal E}$-distributions
corresponding to the color scenarios (I)-(III).
The top (middle) (bottom) part shows
all (`fast': $x > 0.01$)(`slow': $x < 0.01$) clusters, respectively.

\newpage

\begin{center}
{\boldmath $e^+e^- \rightarrow  Z^0 \rightarrow \;hadrons\, , \;\; E_{cm}\;=\;
91\; \mbox{GeV}$}

{\bf Multiplicities: }
\bigskip

\begin{tabular}{l|lllllllllc}
\hline
\hline
                                     &&      &$\;\;\;\;\;\;\;\;\;\;\;\;$&
&$\;\;\;\;\;\;\;\;\;\;\;\;$&   &$\;\;\;\;\;\;\;\;\;\;\;\;$&      &
  \\
Quantity                             && $\;\;\;\;\;\;\;\;$ & I &   & II &  &
III   &&  Experiment \cite{lep,MC}\\
                                            &&      &      &      &      &
&      &&               \\
\hline
\hline
                                            &&      &      &      &      &
&      &&                \\
$\langle n^{ch} \rangle$                    &&   \multicolumn{5}{c}{
$\;\;\;\;\;\;\;\;\;\;\;\;\;\;\;\;\;$21.0 $\;\;$(input)}& &&   $20.95
\;\pm\;0.88$ \\
                                            &&      &      &      &      &
&      &&                \\
$\mu_0$ (GeV)                                    &&  & 0.8  &  & 1.1  &  &  1.5
 &&   $-$ \\
$\langle n_{par} \rangle$                        &&  &24.9  &  & 24.2 &  &19.1
&&   $-$ \\
$\langle n_{had} \rangle$                        &&  &41.3  &  & 42.5 &  &43.2
&&   $-$ \\
$\langle n_{par}\rangle /\langle n_{had}\rangle$ &&  & 0.60 &  & 0.57 &  & 0.44
&&   $-$ \\
                                            &&      &      &      &      &
&      &&               \\
$\langle n^{\pi^\pm}\rangle$                &&  & 16.6&  & 15.8&  & 17.7 &&
17.1 $\pm$ 0.4 \\
$\langle n^{\pi^0}\rangle$                  &&  & 10.2&  & 11.4&  & 10.8 &&
9.9 $\pm$ 0.08\\
$\langle n^{K^\pm} \rangle$                 &&  & 2.79&  & 3.06&  & 2.39 &&
2.42$\pm$ 0.13 \\
$\langle n^{K^0} \rangle$                   &&  & 2.46&  & 2.97&  & 2.03 &&
2.12$\pm$ 0.06 \\
$\langle n^{\rho^\pm} \rangle$              &&  & 2.88&  & 3.09&  & 2.93 &&
   $-$     \\
$\langle n^{\rho^0} \rangle$                &&  & 1.48&  & 1.90&  & 1.67 &&
1.40$\pm$ 0.1 \\
$\langle n^{p} \rangle$                     &&  & 0.72&  & 0.55&  & 0.80 &&
0.92$\pm$ 0.11 \\
$\langle n^{\Lambda^0} \rangle$             &&  & 0.33&  & 0.27&  & 0.39 &&
0.348 $\pm$ 0.013 \\
                                            &&      &     &      &      &
&      &&                \\
\hline
\hline
\end{tabular}

\bigskip

{\Large {\bf Table 1}}
\end{center}

\newpage

\begin{center}
{\boldmath $e^+e^- \rightarrow Z^0 \rightarrow \;hadrons\, , \;\; E_{cm}\;=\;
91\; \mbox{GeV}$}

{\bf Cluster properties: }
\bigskip

\begin{tabular}{l|llllllll}
\hline
\hline
                                     &&      &$\;\;\;\;\;\;\;\;\;\;\;\;$&
&$\;\;\;\;\;\;\;\;\;\;\;\;$&   &$\;\;\;\;\;\;\;\;\;\;\;\;$                \\
Quantity                             && $\;\;\;\;\;\;\;\;$ & I &   & II &  &
III  \\
                                            &&      &      &      &      &
&            \\
\hline
\hline
                                            &&      &      &      &      &
&            \\
$\langle n_{clu} \rangle$                          &&  & 22.6  &  & 21.5&  &
23.2  \\
$\langle n_{par}\rangle /\langle n_{clu} \rangle$  &&  & 1.10 &  & 1.13 &  &
0.82  \\
$\langle n_{clu}\rangle /\langle n_{had} \rangle$  &&  & 0.55 &  & 0.51 &  &
0.53  \\
$\langle R_{clu} \rangle$ (fm)                  &&  & 0.92 &  & 0.89 &  & 0.97
\\
$\langle M_{clu} \rangle$ (GeV)                 &&  & 1.71 &  & 1.86 &  & 1.67
\\
                                            &&      &      &      &      &
&             \\
$g+g \rightarrow C+C$                    &&  & 52.3 \%&  & 37.2 \%&  & 8.4 \%
\\
$g+g \rightarrow C+g$                    &&  &  $-$   &  &  $-$   &  &15.7 \%
\\
$g+g \rightarrow C+g+g$                  &&  &  $-$   &  &  $-$   &  &33.0 \%
\\
                                            &&      &      &      &      &
&             \\
$q+\bar q \rightarrow C+C$               &&  & 11.3 \%&  &  9.9 \%&  & 2.8 \%
\\
$q+\bar q \rightarrow C+g$               &&  &  $-$   &  &  $-$   &  & 3.1 \%
\\
                                            &&      &      &      &      &
&             \\
$q+g  \rightarrow C+q$                   &&  & 36.4 \%&  & 52.9 \%&  &13.0 \%
\\
$q+g  \rightarrow C+q+g$                 &&  &  $-$   &  &  $-$   &  &23.8 \%
\\
                                            &&      &      &      &      &
&            \\
\hline
\hline
\end{tabular}

\bigskip

{\Large {\bf Table 2}}
\end{center}

\newpage

\begin{center}
{\boldmath $e^+e^- \rightarrow W^+W^- \rightarrow \;hadrons\, , \;\;
E_{cm}\;=\; 170\; \mbox{GeV}$}

{\bf Multiplicities and global event properties:}
\bigskip

\begin{tabular}{l|lllllllllc}
\hline
\hline
                                     &&      &$\;\;\;\;\;\;\;\;\;\;\;\;$&
&$\;\;\;\;\;\;\;\;\;\;\;\;$&   &$\;\;\;\;\;\;\;\;\;\;\;\;$&      &
  \\
Quantity                             && $\;\;\;\;\;\;\;\;$ & I &   & II &  &
III  \\
                                            &&      &      &      &      &
&      &&               \\
\hline
\hline
                                            &&      &      &      &      &
&      &&                \\
$\mu_0$ (GeV)                                      &&  & 0.8  &  & 1.1  &  &
1.5 &&   \\
$\langle n_{par} \rangle$                          &&  &37.4&  & 36.8 &  &29.6
&&    \\
$\langle n_{had} \rangle$                          &&  &68.5&  & 70.7&   &72.2
&&    \\
$\langle n_{par}\rangle /\langle n_{had} \rangle$  &&  &0.55&  & 0.52 &  & 0.41
&&    \\
                                           &&      &      &      &      &
&      &&                \\
$\langle n^{ch} \rangle$                    &&  &36.1&  & 37.3 &  &37.5  &&
\\
$\langle n^{ch}\rangle_{|y| < 1}$           &&  & 29.3 &  & 30.3 &  & 30.8 &&
\\
$\langle \;p_\perp^{ch}\;\rangle$ (GeV)     &&  & 1.90 &  & 1.83 &  & 1.75 &&
\\
$\langle \;x_E^{ch}\;\rangle$ ($\times 10^{-2}$)   &&  & 2.85& & 2.74 & & 2.59
&&  \\
$\langle \;1-T \;\rangle$ ($\times 10^{-2}$)  &&  & 6.0&  & 7.1 &  & 6.5 &&  \\
                                           &&      &      &      &      &
&      &&                \\
\hline
\hline
\end{tabular}

\bigskip

{\Large {\bf Table 3}}
\end{center}

\newpage

\begin{center}
{\boldmath $e^+e^- \rightarrow W^+W^- \rightarrow \;hadrons\, , \;\;
E_{cm}\;=\; 170\; \mbox{GeV}$}

{\bf Cluster properties:}
\bigskip

\begin{tabular}{l|llllllll}
\hline
\hline
                                     &&      &$\;\;\;\;\;\;\;\;\;\;\;\;$&
&$\;\;\;\;\;\;\;\;\;\;\;\;$&   &$\;\;\;\;\;\;\;\;\;\;\;\;$                \\
Quantity                             && $\;\;\;\;\;\;\;\;$ & I &   & II &  &
III  \\
                                            &&      &      &      &      &
&            \\

\hline
\hline
                                            &&      &      &      &      &
&            \\
$\langle n_{clu} \rangle$                           &&  & 37.0 &  & 34.8&  &
38.9  \\
$\langle n_{par} \rangle / \langle n_{clu} \rangle$ &&  & 1.01 &  & 1.06 &  &
0.76  \\
$\langle n_{clu} \rangle / \langle n_{had} \rangle$ &&  & 0.54 &  & 0.50 &  &
0.53  \\
$\langle R_{clu} \rangle$ (fm)                  &&  & 0.92 &  & 0.89 &  & 0.97
\\
$\langle M_{clu} \rangle$ (GeV)                 &&  & 1.76 &  & 1.97 &  & 1.62
\\
                                            &&      &      &      &      &
&             \\
$g+g \rightarrow C+C$                    &&  & 47.5 \%&  & 32.6 \%&  & 5.9 \%
\\
$g+g \rightarrow C+g$                    &&  &  $-$   &  &  $-$   &  &15.3 \%
\\
$g+g \rightarrow C+g+g$                   &&  &  $-$   &  &  $-$   &  &25.8 \%
\\
                                            &&      &      &      &      &
&             \\
$q+\bar q \rightarrow C+C$               &&  & 12.8 \%&  & 11.2 \%&  & 2.0 \%
\\
$q+\bar q \rightarrow C+g$               &&  &  $-$   &  &  $-$   &  & 4.4 \%
\\
                                            &&      &      &      &      &
&             \\
$q+g  \rightarrow C+q$                   &&  & 39.7 \%&  & 56.2 \%&  &14.5 \%
\\
$q+g  \rightarrow C+q+g$                 &&  &  $-$   &  &  $-$   &  &32.2 \%
\\
                                            &&      &      &      &      &
&            \\
\hline
\hline
\end{tabular}

\bigskip

{\Large {\bf Table 4}}
\end{center}

\newpage

\begin{center}
{\boldmath $e^+e^- \rightarrow W^+W^- \rightarrow \;hadrons\, , \;\;
E_{cm}\;=\; 170\; \mbox{GeV}$}

{\bf Compilation of mass shifts from {\boldmath $m_W$} reconstruction:}

\bigskip

\begin{tabular}{l|lllllllllc}
\hline
\hline
                                     &&      &$\;\;\;\;\;\;\;\;\;\;\;\;$&
&$\;\;\;\;\;\;\;\;\;\;\;\;$&   &$\;\;\;\;\;\;\;\;\;\;\;\;$&      &
  \\
Quantity  (in GeV)$\;$&& $\;\;\;\;\;\;\;\;$     & I &   & II &  & III   &&  \\
                                            &&      &      &      &      &
&      &&               \\
\hline
\hline
                       \multicolumn{10}{c}{ 1st reconstruction scheme, eq.
(\ref{Cjet1})}  \\
\hline
                                             &&  &     &  &       &  &       &&
   \\
$\langle m_W^{(rec.)} \rangle_{real}$           &&  &80.13 &  & 80.53 &  &80.45
&&    \\
$\langle \Delta m_W \rangle_{real}$          &&  & 0.383 &  & 0.812 &  &0.706
&&    \\
$\sigma (\Delta m_W)_{real}$                 &&  &2.65&  &3.05&  &3.02&&    \\
                                             &&  &     &  &       &  &       &&
   \\
$\langle m_W^{(rec.)} \rangle_{hypo}$           &&  &80.15 &  & 80.55 &  &80.18
&&    \\
$\langle \Delta m_W \rangle_{hypo}$          &&  & 0.396 &  & 0.806 &  &  0.424
&&    \\
$\sigma (\Delta m_W)_{hypo}$                 &&  &2.90&  &3.28&  &3.25&&    \\
                                             &&      &     &      &      &
&      &&                \\
\hline
                       \multicolumn{10}{c}{ 2nd reconstruction scheme, eq.
(\ref{Cjet2})} & \\
\hline
                                             &&  &     &  &       &  &       &&
   \\
$\langle m_W^{(rec.)} \rangle_{real}$           &&  &80.13 &  & 80.30 &  &80.21
&&    \\
$\langle \Delta m_W \rangle_{real}$          &&  & 0.413 &  & 0.899 &  &  0.788
&&    \\
$\sigma (\Delta m_W)_{real}$                 &&  &2.64&  &2.92&  &2.96&&    \\
                                             &&  &     &  &       &  &       &&
   \\
$\langle m_W^{(rec.)} \rangle_{hypo}$           &&  &80.13 &  & 80.28 &  &80.01
&&    \\
$\langle \Delta m_W \rangle_{hypo}$          &&  & 0.422 &  & 0.889 &  &  0.512
&&    \\
$\sigma (\Delta m_W)_{hypo}$                 &&  &2.80&  &3.06&  &3.08&&    \\
                                             &&      &     &      &      &
&      &&                \\

\hline
\hline
\end{tabular}

\bigskip

{\Large {\bf Table 5}}
\end{center}

\vfill

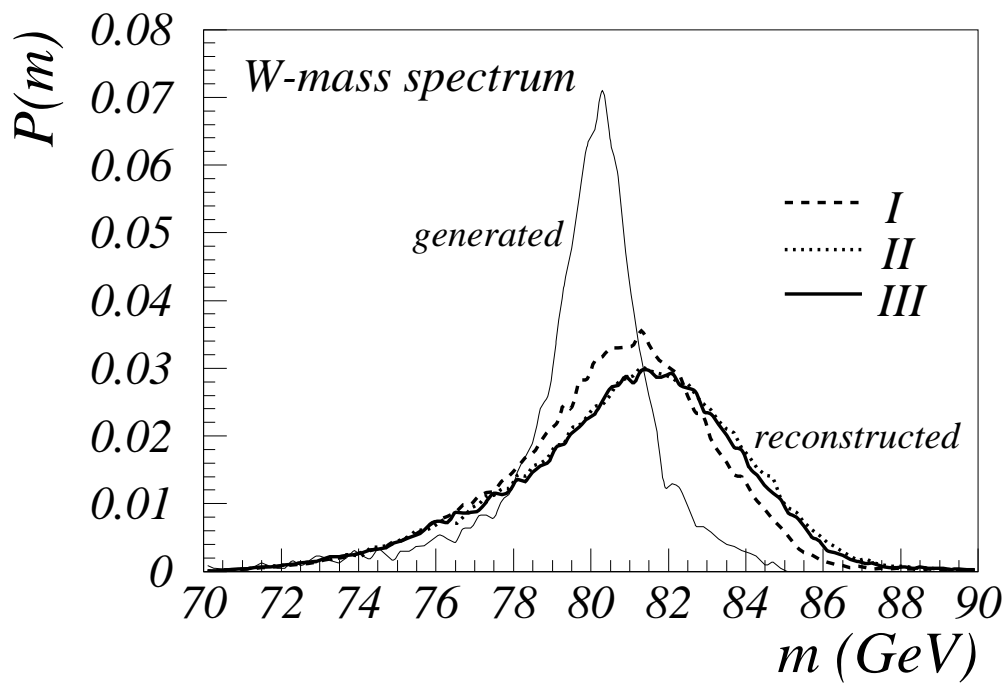
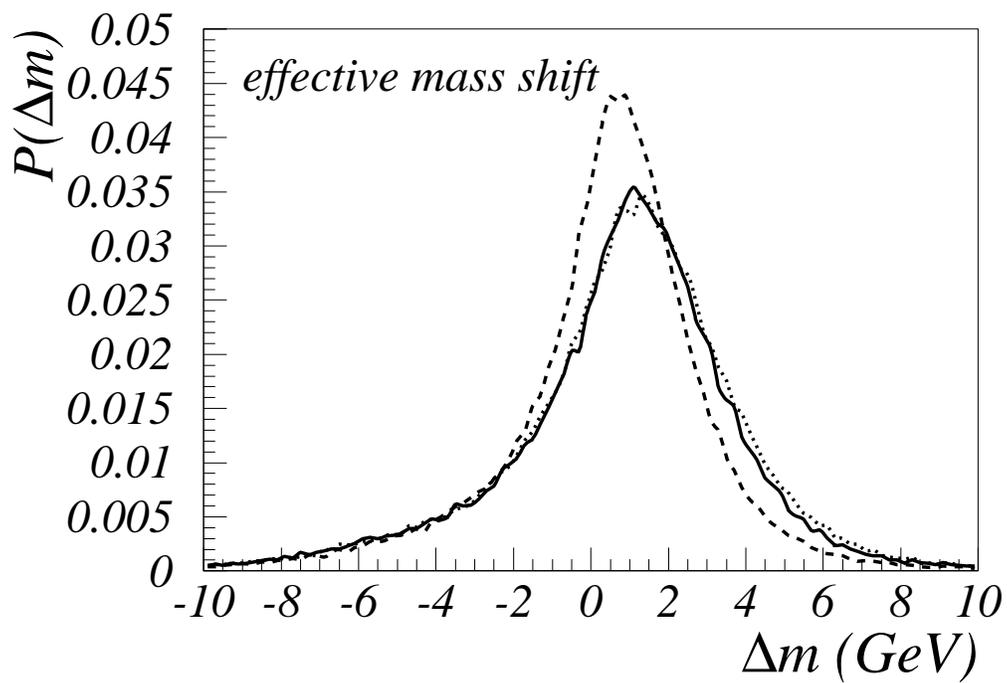
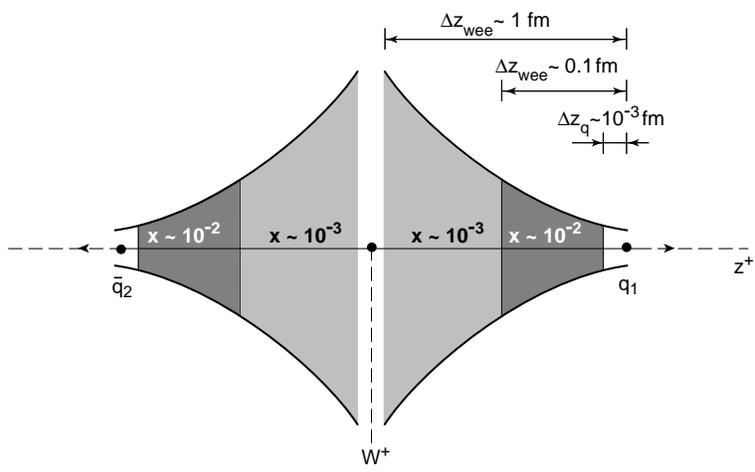
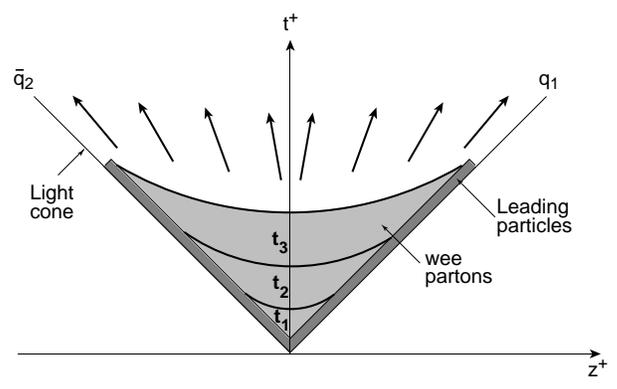
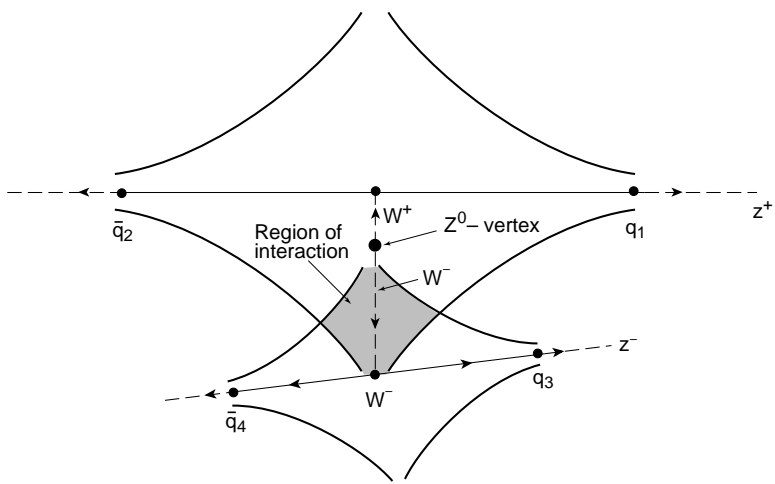
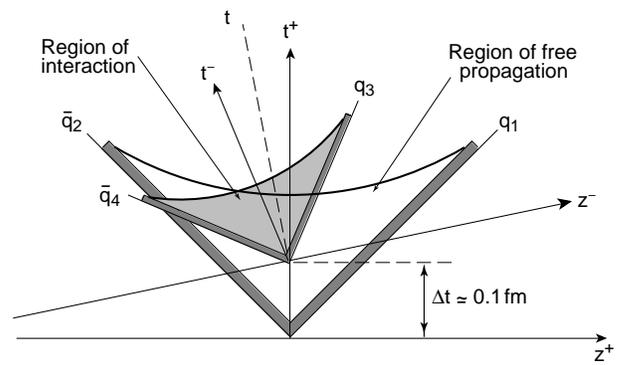
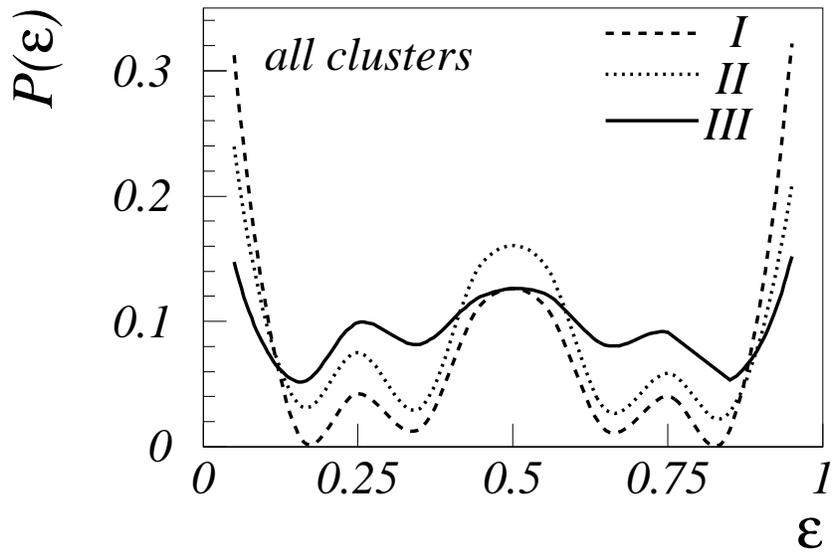
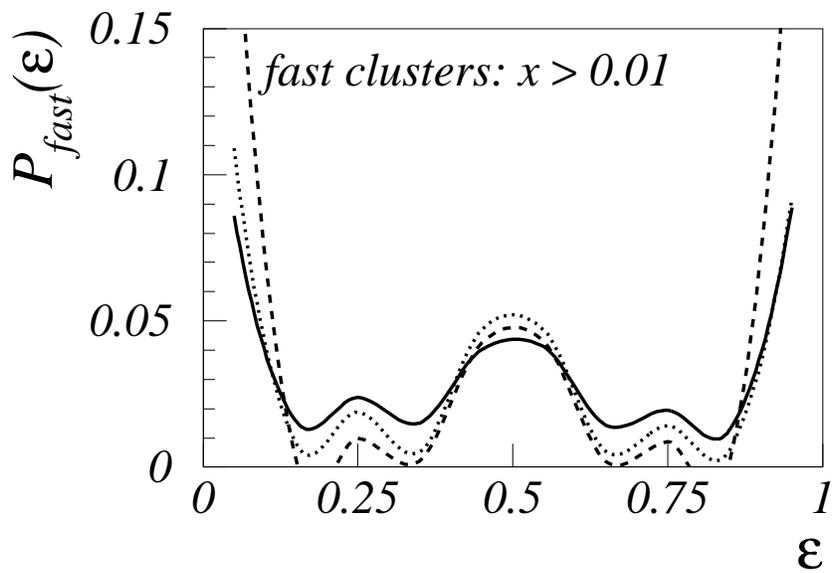
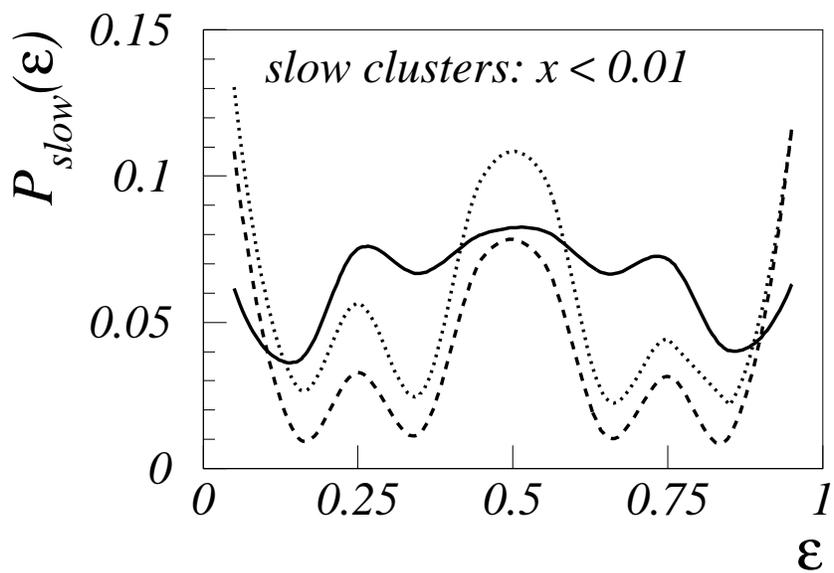
\end{document}